\documentclass[aps,prl,twocolumn,showpacs,superscriptaddress,amsfonts,amsmath,amssymb]{revtex4-1}%
\pdfoutput=1
\usepackage{amsfonts}
\usepackage{color}
\usepackage{amsmath}
\usepackage{amssymb}
\usepackage{graphicx}%
\usepackage{hyperref}
\usepackage{times}
\begin{document}
\title{Unconstrained Capacities of Quantum Key Distribution 
and Entanglement Distillation for 
Pure-Loss Bosonic Broadcast Channels}

%\date{\today}
%
\author{Masahiro Takeoka}

\affiliation{National Institute of Information and Communications Technology,
Koganei, Tokyo 184-8795, Japan}

\author{Kaushik P. Seshadreesan}

\affiliation{Max Planck Institute for the Science of Light, Staudtstr. 2, 91058 Erlangen, Germany}

\author{Mark M. Wilde}

\affiliation{Hearne Institute for Theoretical Physics, Department of Physics
and Astronomy, and Center for Computation and Technology, Louisiana State University, Baton Rouge, Louisiana 70803, USA}

%\pacs{42.50.Dv, 42.50.Ex, 03.67.-a, 03.65.Wj}

%42.50.Dv Quantum state engineering and measurements 
%42.50.Ex Optical implementations of quantum information processing 
%and transfer 
%03.67.-a Quantum information  
%03.65.Wj State reconstruction, quantum tomography  

\begin{abstract}
We consider quantum key distribution (QKD) and entanglement distribution 
using a single-sender multiple-receiver pure-loss bosonic 
broadcast channel. 
We determine the unconstrained 
capacity region for the distillation of bipartite entanglement and 
secret key between the sender and each receiver, whenever they are allowed 
arbitrary public classical communication. 
A practical implication of our result is that 
the capacity region demonstrated drastically improves upon rates  
achievable using a naive time-sharing strategy, which has been employed 
in previously demonstrated network QKD systems. 
We show a simple example of the broadcast QKD protocol 
overcoming the limit of the point-to-point strategy. 
Our result is thus an important step toward opening a new framework of 
network channel-based quantum communication technology. 
\end{abstract}

\maketitle

{\it Introduction}. 
Quantum key distribution (QKD) \cite{BB84,E91} and entanglement distillation
(ED) \cite{BBPSSW95,BDSW96} are two cornerstones of quantum communication.
QKD enables two or more cooperating parties to distill and share 
unconditionally secure, random bit sequences, which could then be used 
for secure classical communication. 
ED, on the other hand, allows them to distill pure 
maximal entanglement from a quantum state shared via a noisy communication
channel, which could then be used to faithfully transfer quantum states
by means of quantum teleportation \cite{BBCJPW93}.
In both protocols, the parties are allowed to perform
(in principle) an unlimited amount of local operations and 
classical communication (LOCC).

Not only have there been a number of theoretical developments, but also quantum communication technologies 
have matured tremendously in recent years.
In particular, QKD has been available commercially for a number of years and
has now expanded to real-world networks \cite{SECOQC09,TOKYO_QKD,ChinaQKD14},
which consist of point-to-point QKD links and trusted nodes.

Another important direction is to go beyond point-to-point links and make
use of network channels. In fact the operation of QKD  
has been proposed for a broadcast channel (single-sender and 
multiple-receiver) \cite{Townsend97} 
and recently experimentally demonstrated for a multiple access channel 
\cite{FDLSYS13} (multiple-sender and single-receiver).  
In \cite{FDLSYS13}, the developed system is based on conventional optical-access network protocols 
(passive optical network), in which the link between each sender 
and receiver is essentially point-to-point quantum communication 
and multiple users share the channel, each having a given  amount of time to use it. 
This \textit{time-sharing protocol} has a strong limit on the rate of key that can be generated
among the parties: when one sender and one receiver 
use the channel most of the time, the key or entanglement rates 
for the other users decrease. 
Then a natural question arises. Is this a fundamental trade-off 
limit or can we do better than the time-sharing limit?

In this paper, we answer this question affirmatively by 
establishing the unconstrained capacity region of 
a pure-loss bosonic broadcast channel, when used for the distillation of 
bipartite entanglement and secret key between the sender and each receiver, along with the 
assistance of unlimited LOCC 
\footnote{Here ``unconstrained capacity'' means the capacity without any energy constraint on the input quantum states.}. 
Even though communication tasks in various network scenarios 
have been examined \cite{AS98,Winter01,GSE07,YHD08,YHD11,DHL10}, 
there has been limited work on the capacity of entanglement and secret key 
distillation assisted by unlimited LOCC. 
Only recently in \cite{STW16} 
were outer bounds on the achievable rates established for
multipartite secret-key agreement and entanglement generation between
any subset of the users of a general single-sender $m$-receiver quantum
broadcast channel (QBC) (for any $m\geq1$), when assisted by unlimited
LOCC between all the users. The main idea was to employ
multipartite generalizations of the squashed entanglement
\cite{AHS08,YHHHOS09} and the methods of \cite{TGW14Nat,TGW14IEEE}.

We break the proof of the capacity region into two parts. The upper bound (converse) is 
established by combining the method in \cite{STW16} and 
the point-to-point upper bound based on relative entropy of entanglement 
\cite{PLOB15,WTB17}, first discussed in \cite{PLOB15} and rigorously proven in \cite{WTB17}.
%\footnote{We note here that versions 1 through 6 of \cite{PLOB15} have contained an incomplete proof of the relative entropy-based bounds for secret key distillation. Looking at (S26) therein (in v6), the parameter $d$ is equal to infinity due to unbounded shield systems and thus the upper bound given in
%\cite{PLOB15} is a trivial bound of $+\infty$. A complete and rigorous proof was thereafter given in \cite{WTB17}. }.
The lower bound (achievability) is proved by employing 
the quantum state merging protocol \cite{HOW05,HOW07}.
Our result clearly shows that the rate region considerably improves upon 
the time-sharing limit, and at the same time, it  proves that this is 
the fundamental limit that cannot be overcome within the same framework. 
Moreover, we do not leave this result as a purely theoretical development, but we also consider 
the possible implementation of a QKD protocol overcoming 
the limit by simple point-to-point protocols for an optical broadcast channel. 
Our result is thus an important step toward the opening of a new framework of 
network channel-based quantum communication technology.

{\it LOCC-assisted distillation in a linear-optical network.} 
We consider the following general entanglement and 
secret-key distillation protocol which uses a quantum broadcast channel \cite{STW16}. 
The sender $A$ prepares some quantum systems in an initial quantum state 
 and successively 
sends some of these systems to the receivers $B_1$, $B_2$, $\ldots, B_m$ 
by interleaving $n$ channel uses of the 1-to-$m$ broadcast channel 
with rounds of LOCC. 
The goal of the protocol is to distill bipartite maximally 
entangled states $\Phi_{AB_i}$ and private states $\gamma_{AB_i}$ 
(equivalently, a secret key \cite{HHHO05,HHHO09}). 
After each channel use, they can perform an arbitrary number of rounds of LOCC 
(in any direction with any number of parties). The quantity 
$E_{AB_i}$ denotes the rate of entanglement that can be established between $A$ and $B_i$ 
(i.e., the logarithm of the Schmidt rank of $\Phi_{AB_i}$ 
normalized by the number of channel uses)
and $K_{AB_i}$ denotes the rate of secret key that can be established between $A$ and $B_i$ 
(i.e., the number of secret-key bits in $\gamma_{AB_i}$ 
normalized by the number of channel uses). The parameter $\varepsilon\in(0,1)$ is such that the fidelity \cite{Uhl76} between the ideal state at the end of the protocol and the actual state is not smaller than $1-\varepsilon$.
The protocol considered here is similar to the one 
described in \cite{STW16}, except that here the goal is \textit{not} to establish 
 bipartite entanglement or key among the receivers or 
 multipartite entanglement or key among more than two parties. 
A rate tuple ($E_{AB_1},\ldots ,E_{AB_m}$, $K_{AB_1}, \ldots, K_{AB_m}$) 
is achievable if for all $\varepsilon \in (0,1)$ and sufficiently large $n$, 
there exists an ($n$, $E_{AB_1},\ldots ,E_{AB_m}$, $K_{AB_1}, \ldots, K_{AB_m}$, $\varepsilon$) 
protocol of the above form. The capacity region is the closure of the set of all achievable rates.

The quantum channel we consider here is a general $1$-to-$m$ bosonic 
broadcast channel $\mathcal{L}_{A' \to B_1 \cdots B_m}$ 
consisting of passive linear optical elements (beam splitters and phase shifters) 
 \cite{G08thesis}. An isometric extension of the channel (see, e.g., \cite{Wil15book}), denoted by 
$U^{\mathcal{L}}$, is then given by an $l$-input $l$-output linear optical 
unitary transformation (see Fig.~1(a)). 
For $U^{\mathcal{L}}$, one of the inputs is the sender $A$'s input and the others are 
prepared as vacuum states. Also, $m$ of the outputs ($m \le l$) are 
given to the legitimate receivers $\{B_1, \ldots , B_m\}$ (one per receiver), and 
the rest of the outputs are for the environment, which we allow the eavesdropper to access during the protocol. The model is thus such that all of the light that does not make it to the legitimate receivers is given to the eavesdropper.
Let $\{\eta_{B_1} , \ldots , \eta_{B_m} \}$ be a set of power transmittances 
from the sender to the respective receivers. Each $\eta_{B_i}$ is non-negative and 
$\sum_{i=1}^{m} \eta_{B_i} \le 1$. 
Let $\mathcal{B} = \{B_1, \ldots ,B_m \}$, 
let $\mathcal{T} \subseteq \mathcal{B}$, 
and let $\overline{\mathcal{T}}$ denote the complement of the set $\mathcal{T}$. 
Then our main result is as follows: 

{\it Theorem 1:} The LOCC-assisted unconstrained capacity region of 
the pure-loss bosonic QBC $\mathcal{L}_{A' \to B_1 \cdots B_m}$ is given by 
%%%%%%%%%%%%%%%%%%%%%%%%%%%
\begin{equation}
\label{eq:capacity_pureloss_m-receiver}
\sum_{B_i \in \mathcal{T}} E_{AB_i} + K_{AB_i} \le 
\log_2 \!\left( 
[1-\eta_{\overline{\mathcal{T}}}]/
[1-\eta_{\mathcal{B}}] 
\right) ,
\end{equation}
%%%%%%%%%%%%%%%%%%%%%%%%%%%
for all non-empty $\mathcal{T} \subseteq \mathcal{B}$, 
where $\eta_\mathcal{B} = \sum_{i=1}^m \eta_{B_i}$ and 
$\eta_{\overline{\mathcal{T}}} 
= \sum_{B_i \in \overline{\mathcal{T}}} \eta_{B_i}$.

%%%%%%%%%%%%%%%%%%%%%%%%%%%%%%%%%%%%%%%%%%%%
\begin{figure}
\begin{center}
\includegraphics[width=90mm]{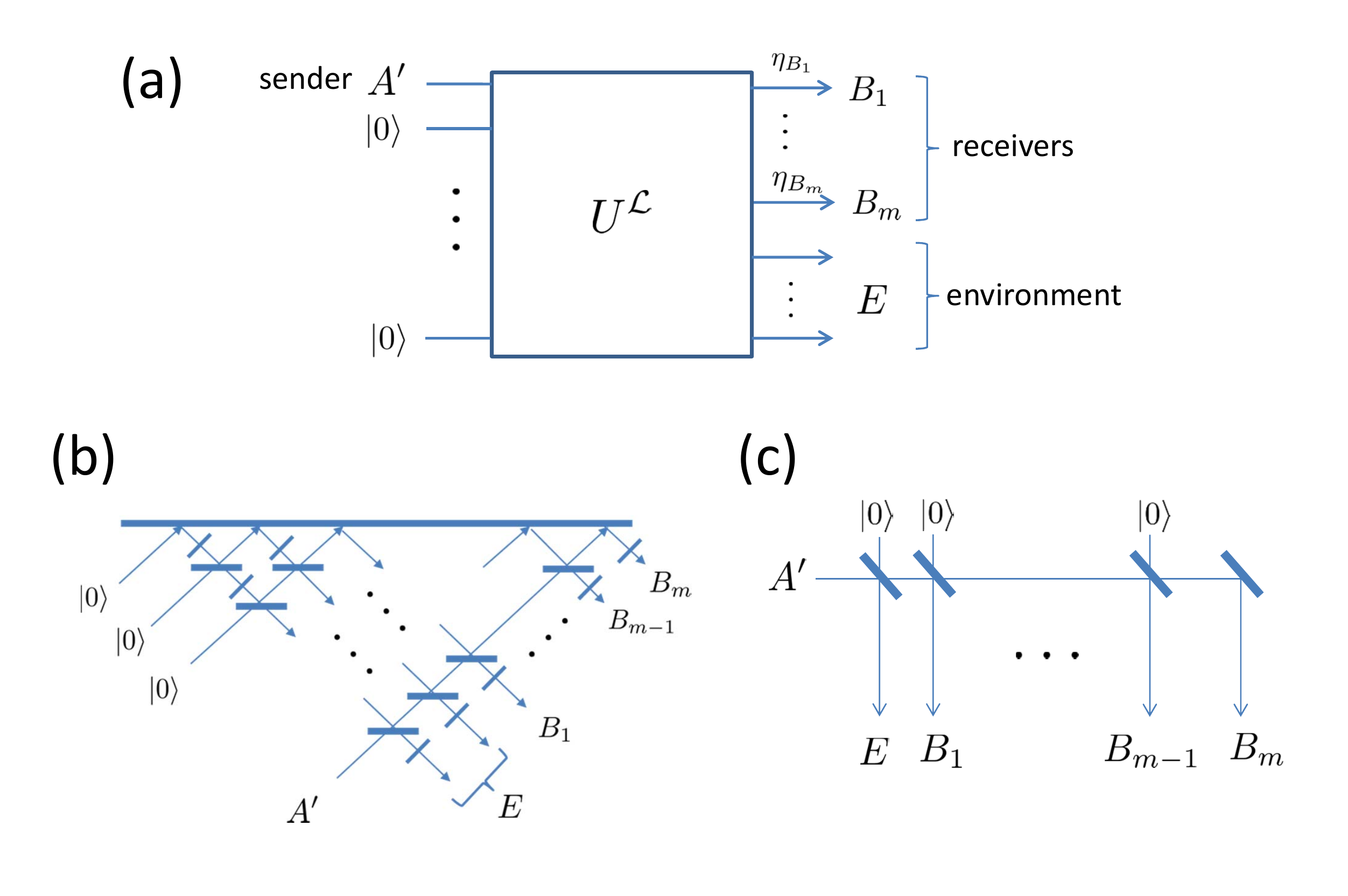}   %
\caption{\label{fig1}
(a) Single-sender $m$-receiver pure-loss linear optics 
quantum broadcast channel. $U^\mathcal{L}$ is 
a unitary operator of an arbitrary linear optics circuit. 
(b), (c) Reductions to an equivalent channel. 
}
\end{center}
\end{figure}
%%%%%%%%%%%%%%%%%%%%%%%%%%%%%%%%%%%%%%%%%%%%

The proof of Theorem~1 consists of three steps: 

{\it (1) Decomposition of $U^{\mathcal{L}}$}. 
First we argue that $U^{\mathcal{L}}$ can be rewritten as 
an equivalent and simpler QBC (see \cite{G08thesis} as well for the reduction outlined here). 
The isometric extension $U^{\mathcal{L}}$ can be represented by 
an $l \times l$ unitary matrix describing the input-output relation 
of a set of annihilation operators $\{\hat{a}_1, \ldots \hat{a}_l\}$ 
for $l$ input modes. 
In \cite{RZBB94}, it was shown that any such $l \times l$ unitary 
matrix can be decomposed as a sequence of 
$2 \times 2$ matrices, each realized by a beam splitter and phase shifters 
combining any two of the $l$ modes (see Fig.~1(b), which contains at most $_l C_2$  beam splitters). 
Now recall that all the inputs except for the sender's are prepared as vacuum states. 
Then we can remove all the beam splitters that have both inputs set to vacuum states because their outputs are vacuum states as well. 
In addition, by grouping together all of the eavesdropper's modes, 
the channel is simplified to just a sequence of $m$ beam splitters 
(Fig.~1(c)). 
In what follows, we consider this simplified, equivalent channel model.

{\it (2) Achievability part}. 
To achieve the rate region in (\ref{eq:capacity_pureloss_m-receiver}), 
we consider a distillation protocol which employs quantum state merging. 
State merging was introduced in \cite{HOW05,HOW07} and provides 
an operational meaning for the conditional quantum entropy. 
For a state $\rho_{AB}$, its conditional quantum entropy is defined as 
$H(A|B)_\rho = H(AB)_\rho - H(B)_\rho$ where $H(AB)_\rho$ and $H(B)_\rho$ 
are the quantum entropies of $\rho_{AB}$ and its marginal $\rho_B$, 
respectively. 
For many copies of $\rho_{AB}$ shared between Alice and Bob, 
$H(A|B)_\rho$ is the optimal rate at which two-qubit maximally entangled  states need
to be consumed to transfer Alice's systems to Bob's side via LOCC. 
If $H(A|B)_\rho$ is negative, the result is that after transferring Alice's 
systems, they can gain (i.e., distill) entanglement at rate $-H(A|B)_\rho$. 
State merging also yields a quantum analog of the Slepian-Wolf theorem 
concerning classical distributed compression and has been applied to the QBC in \cite{YHD11,DHL10}.

Here we consider the following alternative state-merging-based protocol. 
Alice first prepares $n$ copies of a two-mode squeezed vacuum (TMSV) state, defined as
%%%%%%%%%%%%%%%%%%%%%%%%%%%
\begin{equation}
\label{eq:TMSV}
|\Psi(N_S)\rangle_{AA'} = \sum_{m=0}^{\infty} \sqrt{\lambda_m(N_S)} 
|m\rangle_A |m\rangle_{A'},
\end{equation}
%%%%%%%%%%%%%%%%%%%%%%%%%%%
where $|m\rangle$ is an $m$-photon state, 
$
\lambda_m(N_S) = N_S^m/\left(N_S+1\right)^{m+1}, 
$
 and $N_S$ is the average photon number of one mode of the state. 
She sends system $A'$ to $B_1 \cdots B_m$ through 
the broadcast channel in Fig.~1(a). 
After $n$ uses of the channel, they share 
$n$ copies of the state 
$\phi_{AB_1 \cdots B_m} = \mathcal{L}_{A' \to B_1 \cdots B_m} 
(|\Psi(N_S)\rangle \langle \Psi(N_S) | _{AA'})$.

Then by using $\phi_{AB_1 \cdots B_m}^{\otimes n}$, 
they perform state merging to establish entanglement. 
More precisely, all the receivers successively transfer their systems back to Alice by LOCC and at the same time generate entanglement with her in the process
(similar to reverse reconciliation in the point-to-point scenario). 
This can be accomplished by applying the point-to-point state merging protocol 
successively \cite{HOW05,HOW07}. 

Then we obtain the achievable rate region as 
%%%%%%%%%%%%%%%%%%%%%%%%%%%
\begin{equation}
\label{eq:SM_LB_m-receiver}
\sum_{B_i \in \mathcal{T}} E_{AB_i} \le 
- H(\mathcal{T}|A \overline{\mathcal{T}})_\phi ,
\end{equation}
%%%%%%%%%%%%%%%%%%%%%%%%%%%
where 
%%%%%%%%%%%%%%%%%%%%%%%%%%%
$
%\label{eq:phi_m}
\phi_{AB_1 \cdots B_m} = 
\mathcal{L}_{A' \to B_1 \cdots B_m} 
(|\Psi(N_S)\rangle\langle\Psi(N_S)|_{AA'}) 
$.
%%%%%%%%%%%%%%%%%%%%%%%%%%%
The right-hand side of the inequalities in \eqref{eq:SM_LB_m-receiver} can be explicitly calculated. 
Recall that the marginal of the TMSV 
$\Psi_{A'}(N_S) = {\rm Tr}_A [ |\Psi(N_S)\rangle
\langle \Psi(N_S) |_{AA'} ]$ 
is a thermal state with mean photon number $N_S$. 
Its entropy is equal to $H(A')_{\Psi} = g(N_S)$, where 
$g(x) = (x + 1) \log_2 (x+1) - x \log_2 x$. 
Also a pure-loss channel with transmittance $\eta$ 
maps a thermal state to another thermal state 
with reduced average photon number. 
Then the right-hand side of (\ref{eq:SM_LB_m-receiver}) is 
calculated as 
%%%%%%%%%%%%%%%%%%%%%%%%%%%
\begin{align}
\label{eq:-H(T|ATbar)}
-H(\mathcal{T}|A \overline{\mathcal{T}})_\phi & =  
H(A \overline{\mathcal{T}})_\phi - 
H(A \mathcal{T} \overline{\mathcal{T}})_\phi 
\nonumber\\ & =  
H(\mathcal{T}E)_\phi - H(E)_\phi 
\nonumber\\ & =  
g((1-\eta_{\overline{\mathcal{T}}})N_S) - 
g((1-\eta_{\mathcal{B}})N_S).
\nonumber
\end{align}
%%%%%%%%%%%%%%%%%%%%%%%%%%%
By taking $N_S \to \infty$ in the last line above, the limit is equal to 
$\log_2 ([1-\eta_{\overline{\mathcal{T}}}]/[1-\eta_{\mathcal{B}}])$. 
Since one ebit of entanglement can generate one private bit of key, 
we can replace $E_{AB_i}$ with $E_{AB_i}+K_{AB_i}$, which completes 
the achievability part.

{\it (3) Converse part}. 
The converse relies upon several tools and is given 
in terms of the one-shot variant \cite{BD11} of the relative entropy of entanglement (REE) \cite{VP98}.  
The $\varepsilon$-REE for a quantum state $\rho_{AB}$ is defined by 
%%%%%%%%%%%%%%%%%%%%%%%%%%%
\begin{equation}
\label{eq:REE}
E^{\varepsilon}_R(A;B)_\rho = 
\inf_{\sigma_{AB}\in\textrm{SEP}}D_H^{\varepsilon}(\rho_{AB}\Vert \sigma_{AB} ) ,
\end{equation}
%%%%%%%%%%%%%%%%%%%%%%%%%%%
where $D_H^{\varepsilon}(\rho\Vert\sigma) =
-\log_2 \inf_{0\leq \Lambda \leq I, \operatorname{Tr}[\Lambda\rho]\geq 1-\varepsilon}\operatorname{Tr}[\Lambda \sigma]$ 
is the hypothesis testing quantum relative entropy \cite{HP91,BD10,WR12} and SEP denotes the set of  
separable states.
The original LOCC-assisted communication protocol can equivalently be
rewritten by using a teleportation simulation argument  
 \cite[Section~V]{BDSW96} (see also \cite{Mul12}) suitably extended
to continuous-variable bosonic channels \cite{NFC09}. Teleportation simulation in the case of a point-to-point channel 
can be understood as a way of reducing a sequence 
of adaptive protocols involving two-way LOCC to a sequence 
of non-adaptive protocols followed by a final LOCC
 \cite{BDSW96,Mul12}.
For all `teleportation-simulable channels' (more precisely the channels arising
from the action of a generalized teleportation protocol on a bipartite state) that allow for such a reduction, 
an upper bound on the entanglement and secret key agreement capacity 
can be given by the $\varepsilon$-REE \cite{WTB17}, because the $\varepsilon$-REE is an upper bound on the one-shot distillable key
of a bipartite state \cite{WTB17}. 
Furthermore, for pure-loss bosonic channels, one can use a concise formula for the REE identified in \cite{PLOB15}. 
With these techniques, an upper bound on the unconstrained capacity of 
a point-to-point pure-loss channel is given by 
 the REE of the state resulting from sending an infinite-energy TMSV through the channel, explicitly calculated to be $-\log_2 (1-\eta)$ \cite{PLOB15,WTB17}.

Following \cite{WTB17}, suppose that the original protocol generates a state $\omega_{AB_1 \cdots B_m}$ which is $\varepsilon$-close to 
$\tilde{\Phi}_{AB_1 \cdots B_m}$:  
\begin{align}
1- \varepsilon & \leq F( \omega_{AB_1 \cdots B_m} , \tilde{\Phi}_{AB_1 \cdots B_m}) 
  \\
\tilde{\Phi}_{AB_1 \cdots B_m}  & =  
\Phi_{A^{1} B_1^{1}}^{\otimes n E_{AB_1}} \otimes \cdots 
\otimes \Phi_{A^{m} B_m^{1}}^{\otimes n E_{AB_m}} 
\nonumber\\  
& \qquad \otimes 
\gamma_{A^{m+1} B_1^{2}}^{\otimes n K_{AB_1}} \otimes \cdots 
\otimes \gamma_{A^{2m} B_m^{2}}^{\otimes n K_{AB_m}} ,
\label{eq:ideal_state}
\end{align}
%%%%%%%%%%%%%%%%%%%%%%%%%%%
where $A^j$ and $B^j_i$ are subsystems of 
$A$ and $B_i$, respectively.
Since the pure-loss bosonic QBC is covariant with respect to displacement operations (which are the teleportation corrections for bosonic channels \cite{prl1998braunstein}), it is teleportation-simulable \cite{NFC09}. 
Then the original broadcasting protocol described above 
can be replaced by the distillation of $n$ copies of 
$\phi_{AB_1 \cdots B_m} = \mathcal{N}_{A' \to B_1 \cdots B_m} 
(|\Psi(N_S)\rangle\langle \Psi(N_S)|_{AA'})$ 
via a single LOCC. This simulation incurs an additional error that depends on the energy $N_S$ of the state $|\Psi(N_S)\rangle_{AA'}$  and which vanishes in the limit $N_S \to \infty$. 
Let us denote the total error by $\varepsilon(N_S)$ and note that $\lim_{N_S \to \infty} \varepsilon(N_S)= \varepsilon$. Then by using the arguments of \cite{WTB17}, 
we find that 
%%%%%%%%%%%%%%%%%%%%%%%%%%%
\begin{align}
\label{eq:REE_bound_m-receiver}
 \sum_{B_i \in \mathcal{T}} (E_{AB_i}+K_{AB_i}) 
& \le  \frac{1}{n} E_R^{\varepsilon(N_S)}(\mathcal{T}^n;A^n\overline{\mathcal{T}}^n)_{\phi^{\otimes n}} 
%\nonumber\\ & \le  
%E_R(\mathcal{T};A\overline{\mathcal{T}})_{\omega} 
%+ f_\mathcal{T}(n, \varepsilon) 
%\nonumber\\ & \le  
%n E_R(\mathcal{T};A\overline{\mathcal{T}})_{\phi} 
%+ f_\mathcal{T}(n, \varepsilon) ,
\end{align}
%%%%%%%%%%%%%%%%%%%%%%%%%%%
for all $\mathcal{T}$, 
where $E_R^{\varepsilon(N_S)}$ denotes the $\varepsilon$-relative entropy of entanglement  recalled above.

To find an upper bound on  $\frac{1}{n} E_R^{\varepsilon(N_S)}(\mathcal{T}^n;A^n\overline{\mathcal{T}}^n)_{\phi^{\otimes n}} $ for each 
$\mathcal{T}$, we use a calculation from \cite{PLOB15,WTB17} 
for a point-to-point pure-loss bosonic channel with transmittance $\eta$. In \cite{WTB17}, it was shown that the $\varepsilon$-relative entropy of entanglement for a pure-loss channel is bounded from above by
\begin{equation} 
-\log_2(1-\eta) + C(\varepsilon)/n,
\end{equation}
where $C(\varepsilon) = \log_2 6 + 2 \log_2([1+\varepsilon]/[1-\varepsilon])$.
Also it is critical to observe that the order of the beam splitters in 
Fig.~1(c) is reconfigurable by properly commuting the beam splitting 
operators. 
By using this observation and some properties of the TMSV, we  obtain the following upper bound on \eqref{eq:REE_bound_m-receiver}:
%%%%%%%%%%%%%%%%%%%%%%%%%%%
\begin{equation}
\label{eq:REE_m-receiver}
 \log_2\! \left( [1-\eta_{\overline{\mathcal{T}}}]/[1-\eta_{\mathcal{B}}] 
\right) + C(\varepsilon)/n.
\end{equation}
%%%%%%%%%%%%%%%%%%%%%%%%%%%
The converse proof is completed by taking the limit $n\to \infty$. Note that our converse is a {\it strong converse} because there is no need to take the limit $\varepsilon \to 0$ in order to get the upper bound of $\log_2\! \left( \frac{1-\eta_{\overline{\mathcal{T}}}}{1-\eta_{\mathcal{B}}} 
\right)$.
See Supp.~Mat.~1 for more technical details of the calculation.  
Considering that the converse bound coincides with the achievable rate region, 
this completes the proof of Theorem 1.

%Bosonic channels are important in practice as they form a simple model for free-space or fiber-optic communication. Establishing the capacity is one of the goals in quantum Shannon theory and our result (Theorem 1) contributes to this goal in LOCC-assisted entanglement and secret key distillation in quantum network channels. In addition, it also has strong implications for practical quantum information technology, particularly in QKD. The previous proposal of a broadcast QKD \cite{Townsend97} and demonstration of a multiple-access QKD \cite{FDLSYS13} considered only time-sharing strategies, in which each sender and receiver essentially implement  point-to-point QKD and share the channel in time for each combination of sender and receiver. An important question is whether it is possible to improve upon this simple time-sharing strategy. 

{\it Discussion.} 
The simplest pure-loss broadcast channel is a 1-to-2 broadcast channel 
with one sender, Alice, and two receivers, Bob and Charlie. 
The capacity region implied by Theorem~1 is explicitly given by 
%%%%%%%%%%%%%%%%%%%%%%%%%%%
\begin{align}
\label{eq:capacity_1-to-2QBC1}
& E_{AB} + K_{AB}  \le  
\log_2 \!\left([1-\eta_C]/[1-\eta_B-\eta_C] \right), 
\\
\label{eq:capacity_1-to-2QBC2}
& E_{AC} + K_{AC}  \le  
\log_2 \!\left([1-\eta_B]/[1-\eta_B-\eta_C] \right), 
\\
\label{eq:capacity_1-to-2QBC3}
& \!\!\!\!E_{AB} + K_{AB} + E_{AC} + K_{AC}  \le -
\log_2 \!\left(1-\eta_B-\eta_C\right), 
\end{align}
%%%%%%%%%%%%%%%%%%%%%%%%%%%
where $\eta_B$ and $\eta_C$ are the transmittances from Alice to 
Bob and Charlie, respectively. 

\begin{figure}
\begin{center}
\includegraphics[width=90mm]{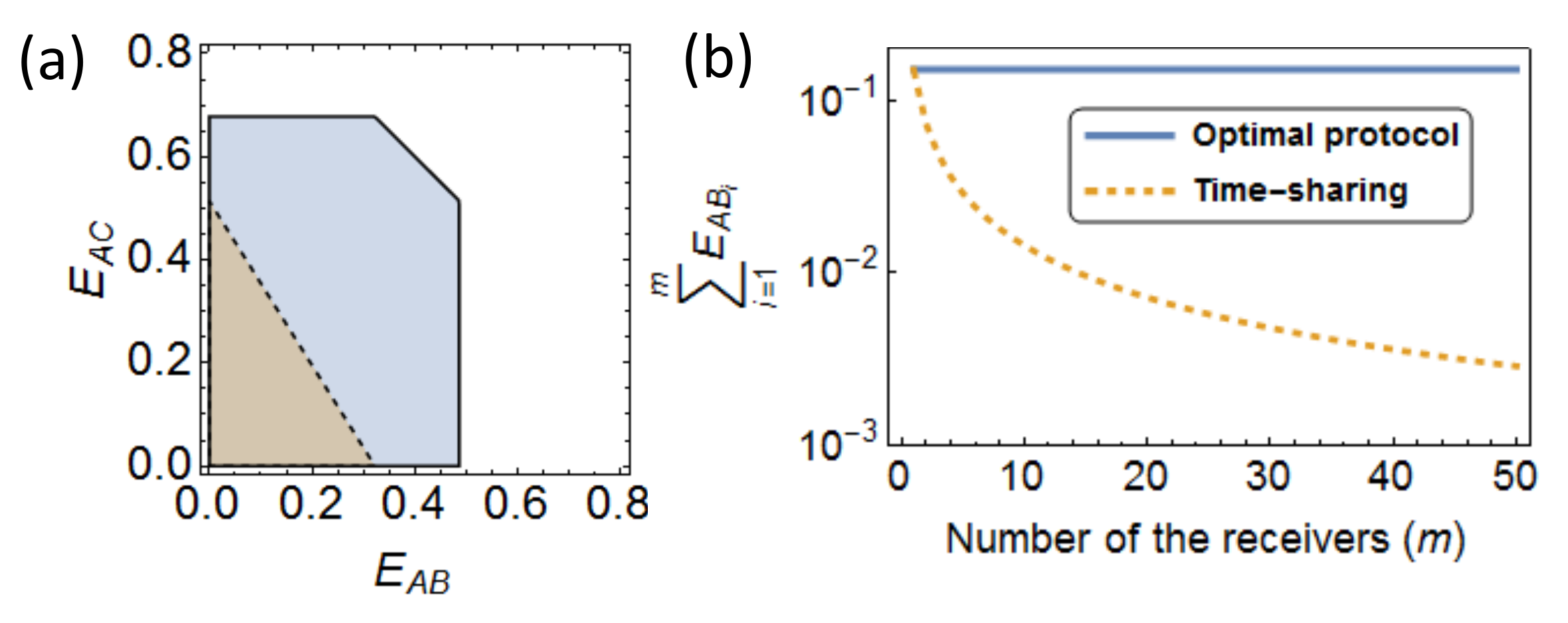}   %
\caption{\label{capacity_region}
Comparison of the LOCC-assisted capacity (solid line) and the time sharing of the point-to-point capacity (dashed line). 
(a) Capacity region for the 1-to-2 QBC with $(\eta_B,\, \eta_C) = (0.2, \, 0.3)$. 
(b) Rate sum comparison for the 1-to-$m$ QBC with $\eta=0.1$. 
}
\end{center}
\end{figure}

It is interesting to compare the capacity with a point-to-point protocol based approach such as time-sharing. 
To do so, we discuss ED and QKD scenarios separately in what follows. 
A naive ED protocol in QBC is the time sharing of the optimal point-to-point protocol
(i.e., they split $n$ uses of the quantum channel into two parts: distill $E_{AB}$ with rate $-\log_2 (1-\eta_B)$ in the first part 
and $E_{AC}$ with rate $-\log_2 (1-\eta_C)$ in the second part). 
Figure \ref{capacity_region}(a) compares the capacity region and the time-sharing strategy. 
The capacity (optimal strategy) clearly outperforms time sharing and the gap is observed 
even on the axes. 
This rate gain originates from the fact that in the optimal strategy, the third party helps the distillation between the other two 
through a sequence of successive state merging \cite{HOW07} (for example, Charlie helps to increase $E_{AB}$ and vice versa). 

The rate gap is more pronounced when we extend this to the $m$-receiver scenario. 
Consider the 1-to-$m$ symmetric pure-loss channel where each receiver 
has equal transmittance $\eta/m$ and the distillation scenario such that 
all receivers achieve the same rate. 
Then the sum of the rates for the optimal protocol based on state merging 
is $-\log_2 (1-\eta)$ whereas that for time sharing of the point-to-point optimal protocol is $-\log_2 (1-\eta/m)$ 
(see Supp.~Mat.~2 for detailed calculations). 
The plots in Fig.~\ref{capacity_region}(b) show a huge gap between 
time sharing and the optimal key distillation strategy. 

%%%%%%%%%%%%%%%%%%%%%%%%%%%%%%%%%%%%%%%%%%%%
%%%%%%%%%%%%%%%%%%%%%%%%%%%%%%%%%%%%%%%%%%%%

Let us turn to the QKD scenario. 
The experimental demonstration in \cite{FDLSYS13} utilizes the time (or frequency) sharing due to 
technical reasons 
\footnote{In this experiment, the multiple access configuration is chosen to reduce the number of single photon detectors, 
which is the most expensive component in their protocol, where the simultaneous point-to-point protocol does not work.}. 
In principle, however, one can overcome the time-sharing limit by simply applying a point-to-point protocol. 
In QKD protocols, the purpose of quantum communication is usually to hold correlated classical data and then the key distillation is performed classically. 
Thus in the QBC scenario, the sender Alice can copy her data and make point-to-point key distillation simultaneously with 
each receiver (a related idea is in \cite{Townsend97}) which can overcome the tradeoff by time sharing. 
The question is then: can we even outperform this no tradeoff rate region?
Here we show that it is possible by describing an explicit example based on a point-to-point continuous variable QKD protocol 
proposed in \cite{GC09} (GC09),  which uses squeezed state and the key distillation based on  reverse reconciliation 
(see Supp.~Mat.~3 for more details). 

In the 1-to-2 QBC setting, the simultaneous operation of the point-to-point GC09 protocol generates  a pair of key rates 
$(K_{AB}, K_{AC}) =  (I(X;Y)-I(Y;C')_\rho,  I(X;Z)-I(Z;B')_\rho)$, where 
$X$, $Y$, and $Z$ are the classical systems representing the classical data shared by Alice, Bob, and Charlie after $n$ uses of the quantum channel, 
$B'$ and $C'$ are the quantum systems for possible eavesdroppers which may contain 
the environment (usually called Eve) and the receiver who is not involved in the key 
(for example, $C'$ includes Charlie).
$I(X;Z)$ and $I(Z;C')_\rho$ denote classical and quantum mutual information, respectively. 
According to the above observation, they can achieve these two key rates simultaneously. 

\begin{figure}
\begin{center}
\includegraphics[width=90mm]{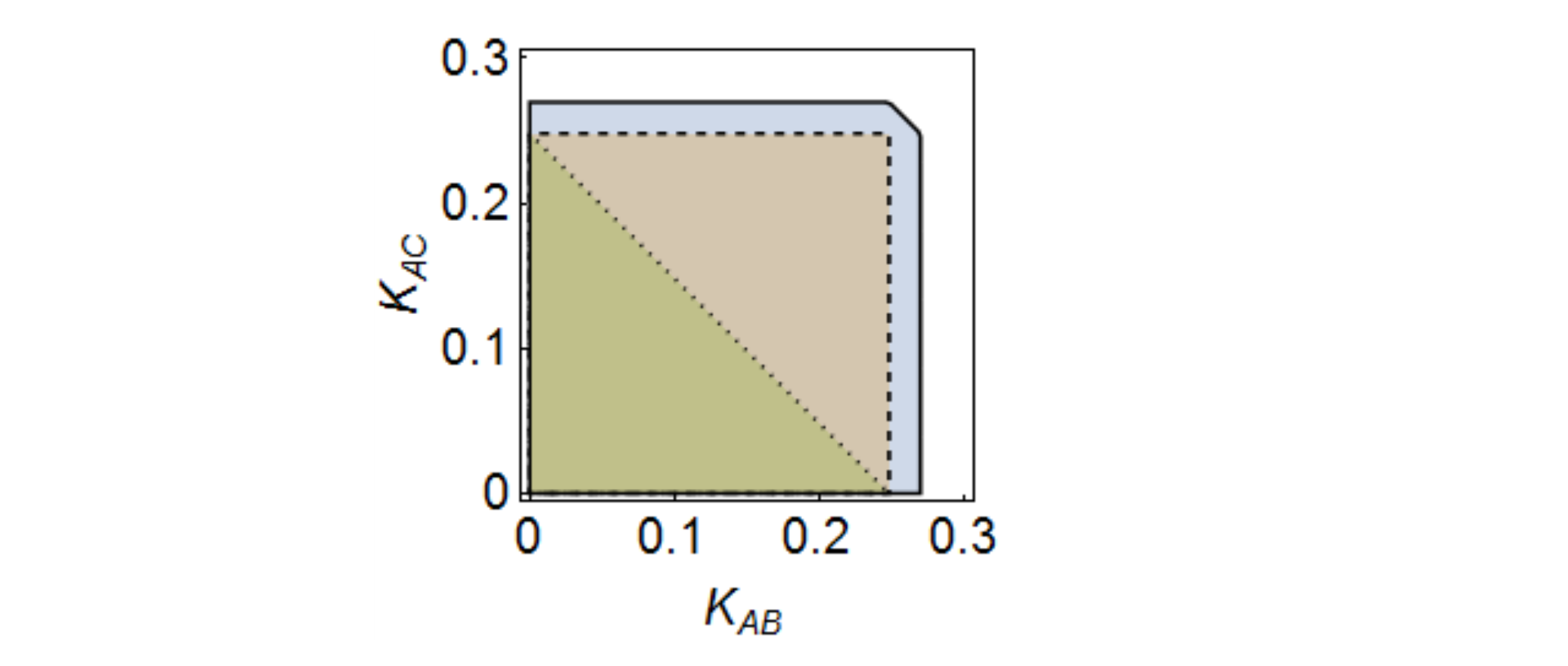}   %
\caption{\label{CVQKD_rate_region} 
Secret key rate region of the GC09 based CVQKD  in a 1-to-2 pure-loss bosonic QBC, with $(\eta_B, \eta_C) = (0.3,0.3)$. 
The BC-CVQKD (solid line), simultaneous application of point-to-point protocol (dashed line), and time sharing 
of the point-to-point protocol (dotted line).
}
\end{center}
\end{figure}

Now we show how to overcome this by using a trick that is similar to the one used in the optimal distillation protocol discussed previously, 
i.e., sequential operation of successive state merging. 
Suppose Alice and Charlie first conduct point-to-point key distillation. 
This operation achieves the key rate $K_{AC} = I(X;Z)-I(Z;B')_\rho$ and also reconstructs 
Charlie's classical system $Z$ at Alice's side. 
After that, Bob distills the key with Alice, where Alice holds $X$ and $Z$ and Bob holds $Y$. 
Thus, they can achieve the key rate $K_{AB} = I(XZ;Y)-I(Y;C')_\rho$, which can be larger than 
that in the simple point-to-point protocol. 
Similarly, by changing the order of distillation, they can achieve the rate pair 
$(K_{AB}, K_{AC})=(I(X;Y)-I(Y;C')_\rho, I(XY;Z)-I(Z;B')_\rho)$. Thus the achievable rate region in this protocol is given by 
time sharing of these two rate pairs. We refer to this protocol as the broadcast-CVQKD (BC-CVQKD). 
Figure \ref{CVQKD_rate_region} shows the key rate region for BC-CVQKD, which outperforms the rate regions for  
the simultaneous point-to-point protocol and the simple time sharing 
(see Supp.~Mat.~3 for an explicit expression of the key rates in a pure-loss QBC). 
Note that the GC09 protocol is originally proposed as a noise-immune CVQKD protocol \cite{GC09} (see also \cite{MULFA12}). 
It is therefore an interesting future work to apply these protocols into a noisy bosonic QBC. 
Also there still remains a huge gap between the key rate region in Fig.~\ref{CVQKD_rate_region} and the capacity region 
in Fig.~\ref{capacity_region}, suggesting that there may exist yet-to-be discovered clever broadcast QKD protocols. 

%%%%%%%%%%%%%%%%%%%%%%%%%%%%%%%%%%%%%%%%%%%%
%%%%%%%%%%%%%%%%%%%%%%%%%%%%%%%%%%%%%%%%%%%%

{\it Conclusion.} 
We have established the unconstrained capacity region of a pure-loss 
bosonic broadcast channel for LOCC-assisted entanglement 
and secret key distillation. 
The channel we considered here is general in the sense that 
it includes any (no-repeater) linear optics network 
as its isometric extension. 
%We prove our result by using the quantum state-merging protocol (for the inner bound) and the relative entropy of entanglement (for the outer  bound). 
It could provide a useful benchmark for the broadcasting 
of entanglement and secret key through such channels. 
Furthermore, our result stimulates practical protocols for 
QKD or entanglement distillation over broadcast channels 
which overcome the time-sharing bound. 
As an example, we discussed a CVQKD protocol based on the proposal in \cite{GC09} 
and show that the BC-CVQKD approach can outperform a simple application of the point-to-point strategy.

%There are interesting problems left open. First, we consider distilling entanglement and a secret key between the sender and each receiver, but it is interesting to include other possibilities, i.e., sharing entanglement and a key between receivers and even  multipartite entanglement (such as GHZ states) and common secret key. A nontrivial outer bound was established in \cite{STW16} for such a scenario. However, its improvement and the construction of a tight inner bound are open. 

%Second, one might attempt to generalize our scenario to other channels. This could be possible for all broadcast channels for which the teleportation simulation argument is applicable to obtain the REE outer bound. Also it is a pressing open question to determine the constrained capacity region (i.e., with a finite-energy constraint). Since the REE bound using the teleportation reduction technique requires infinite energy to realize an ideal teleportation, one needs an alternative approach here, such as the one given in \cite{STW16} in order to obtain tighter outer bounds in some cases. 

%Last but not least, investigation of practical QKD protocols achieving rates beyond those achievable via the simple application of the point-to-point protcol is an interesting and important direction for practical quantum information technology. An example presented here is just a first step toward this direction. For more practical scenario, one might consider more sophisticated protocols that should also be immune to noise in bosonic channels. 

We thank  Saikat Guha, Michael Horodecki, Haoyu Qi, and John Smolin.
MT acknowledges the Open Partnership Joint Projects of
JSPS Bilateral Joint Research Projects and the ImPACT Program of Council 
for Science, Technology, and Innovation, Japan. 
MMW thanks NICT for hosting  him during Dec.~2015 and acknowledges  NSF and ONR.
KPS thanks the Max Planck Society.

%\section*{References}

%\bibliographystyle{unsrt}
\bibliography{Ref}

\begin{thebibliography}{44}%
\makeatletter
\providecommand \@ifxundefined [1]{%
 \@ifx{#1\undefined}
}%
\providecommand \@ifnum [1]{%
 \ifnum #1\expandafter \@firstoftwo
 \else \expandafter \@secondoftwo
 \fi
}%
\providecommand \@ifx [1]{%
 \ifx #1\expandafter \@firstoftwo
 \else \expandafter \@secondoftwo
 \fi
}%
\providecommand \natexlab [1]{#1}%
\providecommand \enquote  [1]{``#1''}%
\providecommand \bibnamefont  [1]{#1}%
\providecommand \bibfnamefont [1]{#1}%
\providecommand \citenamefont [1]{#1}%
\providecommand \href@noop [0]{\@secondoftwo}%
\providecommand \href [0]{\begingroup \@sanitize@url \@href}%
\providecommand \@href[1]{\@@startlink{#1}\@@href}%
\providecommand \@@href[1]{\endgroup#1\@@endlink}%
\providecommand \@sanitize@url [0]{\catcode `\\12\catcode `\$12\catcode
  `\&12\catcode `\#12\catcode `\^12\catcode `\_12\catcode `\%12\relax}%
\providecommand \@@startlink[1]{}%
\providecommand \@@endlink[0]{}%
\providecommand \url  [0]{\begingroup\@sanitize@url \@url }%
\providecommand \@url [1]{\endgroup\@href {#1}{\urlprefix }}%
\providecommand \urlprefix  [0]{URL }%
\providecommand \Eprint [0]{\href }%
\providecommand \doibase [0]{http://dx.doi.org/}%
\providecommand \selectlanguage [0]{\@gobble}%
\providecommand \bibinfo  [0]{\@secondoftwo}%
\providecommand \bibfield  [0]{\@secondoftwo}%
\providecommand \translation [1]{[#1]}%
\providecommand \BibitemOpen [0]{}%
\providecommand \bibitemStop [0]{}%
\providecommand \bibitemNoStop [0]{.\EOS\space}%
\providecommand \EOS [0]{\spacefactor3000\relax}%
\providecommand \BibitemShut  [1]{\csname bibitem#1\endcsname}%
\let\auto@bib@innerbib\@empty
%</preamble>
\bibitem [{\citenamefont {Bennett}\ and\ \citenamefont
  {Brassard}(1984)}]{BB84}%
  \BibitemOpen
  \bibfield  {author} {\bibinfo {author} {\bibfnamefont {C.~H.}\ \bibnamefont
  {Bennett}}\ and\ \bibinfo {author} {\bibfnamefont {G.}~\bibnamefont
  {Brassard}},\ }\href@noop {} {\bibfield  {journal} {\bibinfo  {journal}
  {Proceedings of the IEEE International Conference on Computers, Systems, and
  Signal Processing}\ ,\ \bibinfo {pages} {175}} (\bibinfo {year}
  {1984})}\BibitemShut {NoStop}%
\bibitem [{\citenamefont {Ekert}(1991)}]{E91}%
  \BibitemOpen
  \bibfield  {author} {\bibinfo {author} {\bibfnamefont {A.~K.}\ \bibnamefont
  {Ekert}},\ }\href@noop {} {\bibfield  {journal} {\bibinfo  {journal}
  {Physical Review Letters}\ }\textbf {\bibinfo {volume} {67}},\ \bibinfo
  {pages} {661} (\bibinfo {year} {1991})}\BibitemShut {NoStop}%
\bibitem [{\citenamefont {Bennett}\ \emph
  {et~al.}(1996{\natexlab{a}})\citenamefont {Bennett}, \citenamefont
  {Brassard}, \citenamefont {Popescu}, \citenamefont {Schumacher},
  \citenamefont {Smolin},\ and\ \citenamefont {Wootters}}]{BBPSSW95}%
  \BibitemOpen
  \bibfield  {author} {\bibinfo {author} {\bibfnamefont {C.~H.}\ \bibnamefont
  {Bennett}}, \bibinfo {author} {\bibfnamefont {G.}~\bibnamefont {Brassard}},
  \bibinfo {author} {\bibfnamefont {S.}~\bibnamefont {Popescu}}, \bibinfo
  {author} {\bibfnamefont {B.}~\bibnamefont {Schumacher}}, \bibinfo {author}
  {\bibfnamefont {J.~A.}\ \bibnamefont {Smolin}}, \ and\ \bibinfo {author}
  {\bibfnamefont {W.~K.}\ \bibnamefont {Wootters}},\ }\href@noop {} {\bibfield
  {journal} {\bibinfo  {journal} {Physical Review Letters}\ }\textbf {\bibinfo
  {volume} {76}},\ \bibinfo {pages} {722} (\bibinfo {year}
  {1996}{\natexlab{a}})}\BibitemShut {NoStop}%
\bibitem [{\citenamefont {Bennett}\ \emph
  {et~al.}(1996{\natexlab{b}})\citenamefont {Bennett}, \citenamefont
  {DiVincenzo}, \citenamefont {Smolin},\ and\ \citenamefont
  {Wootters}}]{BDSW96}%
  \BibitemOpen
  \bibfield  {author} {\bibinfo {author} {\bibfnamefont {C.~H.}\ \bibnamefont
  {Bennett}}, \bibinfo {author} {\bibfnamefont {D.~P.}\ \bibnamefont
  {DiVincenzo}}, \bibinfo {author} {\bibfnamefont {J.~A.}\ \bibnamefont
  {Smolin}}, \ and\ \bibinfo {author} {\bibfnamefont {W.~K.}\ \bibnamefont
  {Wootters}},\ }\href@noop {} {\bibfield  {journal} {\bibinfo  {journal}
  {Physical Review A}\ }\textbf {\bibinfo {volume} {54}},\ \bibinfo {pages}
  {3824} (\bibinfo {year} {1996}{\natexlab{b}})},\ \Eprint
  {http://arxiv.org/abs/9604024} {arXiv:9604024 [quant-ph]} \BibitemShut
  {NoStop}%
\bibitem [{\citenamefont {Bennett}\ \emph {et~al.}(1993)\citenamefont
  {Bennett}, \citenamefont {Brassard}, \citenamefont {Cr\'epeau}, \citenamefont
  {Jozsa}, \citenamefont {Peres},\ and\ \citenamefont {Wootters}}]{BBCJPW93}%
  \BibitemOpen
  \bibfield  {author} {\bibinfo {author} {\bibfnamefont {C.~H.}\ \bibnamefont
  {Bennett}}, \bibinfo {author} {\bibfnamefont {G.}~\bibnamefont {Brassard}},
  \bibinfo {author} {\bibfnamefont {C.}~\bibnamefont {Cr\'epeau}}, \bibinfo
  {author} {\bibfnamefont {R.}~\bibnamefont {Jozsa}}, \bibinfo {author}
  {\bibfnamefont {A.}~\bibnamefont {Peres}}, \ and\ \bibinfo {author}
  {\bibfnamefont {W.~K.}\ \bibnamefont {Wootters}},\ }\href@noop {} {\bibfield
  {journal} {\bibinfo  {journal} {Physical Review Letters}\ }\textbf {\bibinfo
  {volume} {70}},\ \bibinfo {pages} {1895} (\bibinfo {year}
  {1993})}\BibitemShut {NoStop}%
\bibitem [{\citenamefont {Peev}\ \emph {et~al.}(2009)\citenamefont {Peev},
  \citenamefont {Pacher},\ and\ \citenamefont {All{\'{e}}aume}}]{SECOQC09}%
  \BibitemOpen
  \bibfield  {author} {\bibinfo {author} {\bibfnamefont {M.}~\bibnamefont
  {Peev}}, \bibinfo {author} {\bibfnamefont {C.}~\bibnamefont {Pacher}}, \ and\
  \bibinfo {author} {\bibfnamefont {R.}~\bibnamefont {All{\'{e}}aume}},\
  }\href@noop {} {\bibfield  {journal} {\bibinfo  {journal} {New Journal of
  Physics}\ }\textbf {\bibinfo {volume} {11}},\ \bibinfo {pages} {075001}
  (\bibinfo {year} {2009})}\BibitemShut {NoStop}%
\bibitem [{\citenamefont {{M. Sasaki et al.}}(2011)}]{TOKYO_QKD}%
  \BibitemOpen
  \bibfield  {author} {\bibinfo {author} {\bibnamefont {{M. Sasaki et al.}}},\
  }\href@noop {} {\bibfield  {journal} {\bibinfo  {journal} {Optics Express}\
  }\textbf {\bibinfo {volume} {19}},\ \bibinfo {pages} {10387} (\bibinfo {year}
  {2011})},\ \bibinfo {note} {arXiv:quant-ph/1103.3566}\BibitemShut {NoStop}%
\bibitem [{\citenamefont {{S. Wang et al.}}(2014)}]{ChinaQKD14}%
  \BibitemOpen
  \bibfield  {author} {\bibinfo {author} {\bibnamefont {{S. Wang et al.}}},\
  }\href {\doibase 10.1364/OE.22.021739} {\bibfield  {journal} {\bibinfo
  {journal} {Optics Express}\ }\textbf {\bibinfo {volume} {22}},\ \bibinfo
  {pages} {21739} (\bibinfo {year} {2014})}\BibitemShut {NoStop}%
\bibitem [{\citenamefont {Townsend}(1997)}]{Townsend97}%
  \BibitemOpen
  \bibfield  {author} {\bibinfo {author} {\bibfnamefont {P.~D.}\ \bibnamefont
  {Townsend}},\ }\href {\doibase 10.1038/385047a0} {\bibfield  {journal}
  {\bibinfo  {journal} {Nature}\ }\textbf {\bibinfo {volume} {385}},\ \bibinfo
  {pages} {47} (\bibinfo {year} {1997})}\BibitemShut {NoStop}%
\bibitem [{\citenamefont {Fr{\"{o}}hlich}\ \emph {et~al.}(2013)\citenamefont
  {Fr{\"{o}}hlich}, \citenamefont {Dynes}, \citenamefont {Lucamarini},
  \citenamefont {Sharpe}, \citenamefont {Yuan},\ and\ \citenamefont
  {Shields}}]{FDLSYS13}%
  \BibitemOpen
  \bibfield  {author} {\bibinfo {author} {\bibfnamefont {B.}~\bibnamefont
  {Fr{\"{o}}hlich}}, \bibinfo {author} {\bibfnamefont {J.~F.}\ \bibnamefont
  {Dynes}}, \bibinfo {author} {\bibfnamefont {M.}~\bibnamefont {Lucamarini}},
  \bibinfo {author} {\bibfnamefont {A.~W.}\ \bibnamefont {Sharpe}}, \bibinfo
  {author} {\bibfnamefont {Z.}~\bibnamefont {Yuan}}, \ and\ \bibinfo {author}
  {\bibfnamefont {A.~J.}\ \bibnamefont {Shields}},\ }\href {\doibase
  10.1038/nature12493} {\bibfield  {journal} {\bibinfo  {journal} {Nature}\
  }\textbf {\bibinfo {volume} {501}},\ \bibinfo {pages} {69} (\bibinfo {year}
  {2013})}\BibitemShut {NoStop}%
\bibitem [{Note1()}]{Note1}%
  \BibitemOpen
  \bibinfo {note} {Here ``unconstrained capacity'' means the capacity without
  any energy constraint on the input quantum states.}\BibitemShut {Stop}%
\bibitem [{\citenamefont {Allahverdyan}\ and\ \citenamefont
  {Saakian}(1998)}]{AS98}%
  \BibitemOpen
  \bibfield  {author} {\bibinfo {author} {\bibfnamefont {A.~E.}\ \bibnamefont
  {Allahverdyan}}\ and\ \bibinfo {author} {\bibfnamefont {D.~B.}\ \bibnamefont
  {Saakian}},\ }\href@noop {} {\  (\bibinfo {year} {1998})},\ \Eprint
  {http://arxiv.org/abs/9805067} {arXiv:9805067 [quant-ph]} \BibitemShut
  {NoStop}%
\bibitem [{\citenamefont {Winter}(2001)}]{Winter01}%
  \BibitemOpen
  \bibfield  {author} {\bibinfo {author} {\bibfnamefont {A.}~\bibnamefont
  {Winter}},\ }\href@noop {} {\bibfield  {journal} {\bibinfo  {journal} {IEEE
  Transactions on Information Theory}\ }\textbf {\bibinfo {volume} {47}},\
  \bibinfo {pages} {3059} (\bibinfo {year} {2001})}\BibitemShut {NoStop}%
\bibitem [{\citenamefont {Guha}\ \emph {et~al.}(2007)\citenamefont {Guha},
  \citenamefont {Shapiro},\ and\ \citenamefont {Erkmen}}]{GSE07}%
  \BibitemOpen
  \bibfield  {author} {\bibinfo {author} {\bibfnamefont {S.}~\bibnamefont
  {Guha}}, \bibinfo {author} {\bibfnamefont {J.}~\bibnamefont {Shapiro}}, \
  and\ \bibinfo {author} {\bibfnamefont {B.}~\bibnamefont {Erkmen}},\
  }\href@noop {} {\bibfield  {journal} {\bibinfo  {journal} {Physical Review
  A}\ }\textbf {\bibinfo {volume} {76}},\ \bibinfo {pages} {032303} (\bibinfo
  {year} {2007})}\BibitemShut {NoStop}%
\bibitem [{\citenamefont {Yard}\ \emph {et~al.}(2008)\citenamefont {Yard},
  \citenamefont {Hayden},\ and\ \citenamefont {Devetak}}]{YHD08}%
  \BibitemOpen
  \bibfield  {author} {\bibinfo {author} {\bibfnamefont {J.}~\bibnamefont
  {Yard}}, \bibinfo {author} {\bibfnamefont {P.}~\bibnamefont {Hayden}}, \ and\
  \bibinfo {author} {\bibfnamefont {I.}~\bibnamefont {Devetak}},\ }\href@noop
  {} {\bibfield  {journal} {\bibinfo  {journal} {IEEE Transactions on
  Information Theory}\ }\textbf {\bibinfo {volume} {54}},\ \bibinfo {pages}
  {3091} (\bibinfo {year} {2008})},\ \Eprint {http://arxiv.org/abs/0501045}
  {arXiv:0501045 [quant-ph]} \BibitemShut {NoStop}%
\bibitem [{\citenamefont {Yard}\ \emph {et~al.}(2011)\citenamefont {Yard},
  \citenamefont {Hayden},\ and\ \citenamefont {Devetak}}]{YHD11}%
  \BibitemOpen
  \bibfield  {author} {\bibinfo {author} {\bibfnamefont {J.}~\bibnamefont
  {Yard}}, \bibinfo {author} {\bibfnamefont {P.}~\bibnamefont {Hayden}}, \ and\
  \bibinfo {author} {\bibfnamefont {I.}~\bibnamefont {Devetak}},\ }\href@noop
  {} {\bibfield  {journal} {\bibinfo  {journal} {IEEE Transactions on
  Information Theory}\ }\textbf {\bibinfo {volume} {57}},\ \bibinfo {pages}
  {7147} (\bibinfo {year} {2011})}\BibitemShut {NoStop}%
\bibitem [{\citenamefont {Dupuis}\ \emph {et~al.}(2010)\citenamefont {Dupuis},
  \citenamefont {Hayden},\ and\ \citenamefont {Li}}]{DHL10}%
  \BibitemOpen
  \bibfield  {author} {\bibinfo {author} {\bibfnamefont {F.}~\bibnamefont
  {Dupuis}}, \bibinfo {author} {\bibfnamefont {P.}~\bibnamefont {Hayden}}, \
  and\ \bibinfo {author} {\bibfnamefont {K.}~\bibnamefont {Li}},\ }\href@noop
  {} {\bibfield  {journal} {\bibinfo  {journal} {IEEE Transactions on
  Information Theory}\ }\textbf {\bibinfo {volume} {56}},\ \bibinfo {pages}
  {2946} (\bibinfo {year} {2010})}\BibitemShut {NoStop}%
\bibitem [{\citenamefont {Seshadreesan}\ \emph {et~al.}(2016)\citenamefont
  {Seshadreesan}, \citenamefont {Takeoka},\ and\ \citenamefont
  {Wilde}}]{STW16}%
  \BibitemOpen
  \bibfield  {author} {\bibinfo {author} {\bibfnamefont {K.~P.}\ \bibnamefont
  {Seshadreesan}}, \bibinfo {author} {\bibfnamefont {M.}~\bibnamefont
  {Takeoka}}, \ and\ \bibinfo {author} {\bibfnamefont {M.~M.}\ \bibnamefont
  {Wilde}},\ }\href {\doibase 10.1109/TIT.2016.2544803} {\bibfield  {journal}
  {\bibinfo  {journal} {IEEE Transactions on Information Theory}\ }\textbf
  {\bibinfo {volume} {62}},\ \bibinfo {pages} {2849} (\bibinfo {year}
  {2016})},\ \Eprint {http://arxiv.org/abs/1503.08139} {arXiv:1503.08139}
  \BibitemShut {NoStop}%
\bibitem [{\citenamefont {Avis}\ \emph {et~al.}(2008)\citenamefont {Avis},
  \citenamefont {Hayden},\ and\ \citenamefont {Savov}}]{AHS08}%
  \BibitemOpen
  \bibfield  {author} {\bibinfo {author} {\bibfnamefont {D.}~\bibnamefont
  {Avis}}, \bibinfo {author} {\bibfnamefont {P.}~\bibnamefont {Hayden}}, \ and\
  \bibinfo {author} {\bibfnamefont {I.}~\bibnamefont {Savov}},\ }\href@noop {}
  {\bibfield  {journal} {\bibinfo  {journal} {Journal of Physics A:
  Mathematical and Theoretical}\ }\textbf {\bibinfo {volume} {41}},\ \bibinfo
  {pages} {115301} (\bibinfo {year} {2008})},\ \bibinfo {note}
  {arXiv:0707.2792}\BibitemShut {NoStop}%
\bibitem [{\citenamefont {Yang}\ \emph {et~al.}(2009)\citenamefont {Yang},
  \citenamefont {Horodecki}, \citenamefont {Horodecki}, \citenamefont
  {Horodecki}, \citenamefont {Oppenheim},\ and\ \citenamefont
  {Song}}]{YHHHOS09}%
  \BibitemOpen
  \bibfield  {author} {\bibinfo {author} {\bibfnamefont {D.}~\bibnamefont
  {Yang}}, \bibinfo {author} {\bibfnamefont {K.}~\bibnamefont {Horodecki}},
  \bibinfo {author} {\bibfnamefont {M.}~\bibnamefont {Horodecki}}, \bibinfo
  {author} {\bibfnamefont {P.}~\bibnamefont {Horodecki}}, \bibinfo {author}
  {\bibfnamefont {J.}~\bibnamefont {Oppenheim}}, \ and\ \bibinfo {author}
  {\bibfnamefont {W.}~\bibnamefont {Song}},\ }\href@noop {} {\bibfield
  {journal} {\bibinfo  {journal} {IEEE Transactions on Information Theory}\
  }\textbf {\bibinfo {volume} {55}},\ \bibinfo {pages} {3375} (\bibinfo {year}
  {2009})}\BibitemShut {NoStop}%
\bibitem [{\citenamefont {Takeoka}\ \emph
  {et~al.}(2014{\natexlab{a}})\citenamefont {Takeoka}, \citenamefont {Guha},\
  and\ \citenamefont {Wilde}}]{TGW14Nat}%
  \BibitemOpen
  \bibfield  {author} {\bibinfo {author} {\bibfnamefont {M.}~\bibnamefont
  {Takeoka}}, \bibinfo {author} {\bibfnamefont {S.}~\bibnamefont {Guha}}, \
  and\ \bibinfo {author} {\bibfnamefont {M.~M.}\ \bibnamefont {Wilde}},\
  }\href@noop {} {\bibfield  {journal} {\bibinfo  {journal} {Nature
  Communications}\ }\textbf {\bibinfo {volume} {5}},\ \bibinfo {pages} {5235}
  (\bibinfo {year} {2014}{\natexlab{a}})}\BibitemShut {NoStop}%
\bibitem [{\citenamefont {Takeoka}\ \emph
  {et~al.}(2014{\natexlab{b}})\citenamefont {Takeoka}, \citenamefont {Guha},\
  and\ \citenamefont {Wilde}}]{TGW14IEEE}%
  \BibitemOpen
  \bibfield  {author} {\bibinfo {author} {\bibfnamefont {M.}~\bibnamefont
  {Takeoka}}, \bibinfo {author} {\bibfnamefont {S.}~\bibnamefont {Guha}}, \
  and\ \bibinfo {author} {\bibfnamefont {M.~M.}\ \bibnamefont {Wilde}},\
  }\href@noop {} {\bibfield  {journal} {\bibinfo  {journal} {IEEE Transactions
  on Information Theory}\ }\textbf {\bibinfo {volume} {60}},\ \bibinfo {pages}
  {4987} (\bibinfo {year} {2014}{\natexlab{b}})},\ \bibinfo {note}
  {arXiv:1310.0129}\BibitemShut {NoStop}%
\bibitem [{\citenamefont {Pirandola}\ \emph {et~al.}(2015)\citenamefont
  {Pirandola}, \citenamefont {Laurenza}, \citenamefont {Ottaviani},\ and\
  \citenamefont {Banchi}}]{PLOB15}%
  \BibitemOpen
  \bibfield  {author} {\bibinfo {author} {\bibfnamefont {S.}~\bibnamefont
  {Pirandola}}, \bibinfo {author} {\bibfnamefont {R.}~\bibnamefont {Laurenza}},
  \bibinfo {author} {\bibfnamefont {C.}~\bibnamefont {Ottaviani}}, \ and\
  \bibinfo {author} {\bibfnamefont {L.}~\bibnamefont {Banchi}},\ }\href@noop {}
  {\  (\bibinfo {year} {2015})},\ \Eprint {http://arxiv.org/abs/1510.08863v5}
  {arXiv:1510.08863v5} \BibitemShut {NoStop}%
\bibitem [{\citenamefont {Wilde}\ \emph {et~al.}(2017)\citenamefont {Wilde},
  \citenamefont {Tomamichel},\ and\ \citenamefont {Berta}}]{WTB17}%
  \BibitemOpen
  \bibfield  {author} {\bibinfo {author} {\bibfnamefont {M.~M.}\ \bibnamefont
  {Wilde}}, \bibinfo {author} {\bibfnamefont {M.}~\bibnamefont {Tomamichel}}, \
  and\ \bibinfo {author} {\bibfnamefont {M.}~\bibnamefont {Berta}},\ }\href
  {\doibase 10.1109/TIT.2017.2648825} {\bibfield  {journal} {\bibinfo
  {journal} {IEEE Transactions on Information Theory}\ }\textbf {\bibinfo
  {volume} {63}},\ \bibinfo {pages} {1792} (\bibinfo {year} {2017})},\ \Eprint
  {http://arxiv.org/abs/1602.08898} {arXiv:1602.08898} \BibitemShut {NoStop}%
\bibitem [{\citenamefont {Horodecki}\ \emph
  {et~al.}(2005{\natexlab{a}})\citenamefont {Horodecki}, \citenamefont
  {Oppenheim},\ and\ \citenamefont {Winter}}]{HOW05}%
  \BibitemOpen
  \bibfield  {author} {\bibinfo {author} {\bibfnamefont {M.}~\bibnamefont
  {Horodecki}}, \bibinfo {author} {\bibfnamefont {J.}~\bibnamefont
  {Oppenheim}}, \ and\ \bibinfo {author} {\bibfnamefont {A.}~\bibnamefont
  {Winter}},\ }\href@noop {} {\bibfield  {journal} {\bibinfo  {journal}
  {Nature}\ }\textbf {\bibinfo {volume} {436}},\ \bibinfo {pages} {673}
  (\bibinfo {year} {2005}{\natexlab{a}})}\BibitemShut {NoStop}%
\bibitem [{\citenamefont {Horodecki}\ \emph {et~al.}(2007)\citenamefont
  {Horodecki}, \citenamefont {Oppenheim},\ and\ \citenamefont
  {Winter}}]{HOW07}%
  \BibitemOpen
  \bibfield  {author} {\bibinfo {author} {\bibfnamefont {M.}~\bibnamefont
  {Horodecki}}, \bibinfo {author} {\bibfnamefont {J.}~\bibnamefont
  {Oppenheim}}, \ and\ \bibinfo {author} {\bibfnamefont {A.}~\bibnamefont
  {Winter}},\ }\href@noop {} {\bibfield  {journal} {\bibinfo  {journal}
  {Communications in Mathematical Physics}\ }\textbf {\bibinfo {volume}
  {136}},\ \bibinfo {pages} {107} (\bibinfo {year} {2007})}\BibitemShut
  {NoStop}%
\bibitem [{\citenamefont {Horodecki}\ \emph
  {et~al.}(2005{\natexlab{b}})\citenamefont {Horodecki}, \citenamefont
  {Horodecki}, \citenamefont {Horodecki},\ and\ \citenamefont
  {Oppenheim}}]{HHHO05}%
  \BibitemOpen
  \bibfield  {author} {\bibinfo {author} {\bibfnamefont {K.}~\bibnamefont
  {Horodecki}}, \bibinfo {author} {\bibfnamefont {M.}~\bibnamefont
  {Horodecki}}, \bibinfo {author} {\bibfnamefont {P.}~\bibnamefont
  {Horodecki}}, \ and\ \bibinfo {author} {\bibfnamefont {J.}~\bibnamefont
  {Oppenheim}},\ }\href@noop {} {\bibfield  {journal} {\bibinfo  {journal}
  {Physical Review Letters}\ }\textbf {\bibinfo {volume} {94}},\ \bibinfo
  {pages} {160502} (\bibinfo {year} {2005}{\natexlab{b}})},\ \bibinfo {note}
  {arXiv:quant-ph/0309110}\BibitemShut {NoStop}%
\bibitem [{\citenamefont {Horodecki}\ \emph {et~al.}(2009)\citenamefont
  {Horodecki}, \citenamefont {Horodecki}, \citenamefont {Horodecki},\ and\
  \citenamefont {Oppenheim}}]{HHHO09}%
  \BibitemOpen
  \bibfield  {author} {\bibinfo {author} {\bibfnamefont {K.}~\bibnamefont
  {Horodecki}}, \bibinfo {author} {\bibfnamefont {M.}~\bibnamefont
  {Horodecki}}, \bibinfo {author} {\bibfnamefont {P.}~\bibnamefont
  {Horodecki}}, \ and\ \bibinfo {author} {\bibfnamefont {J.}~\bibnamefont
  {Oppenheim}},\ }\href@noop {} {\bibfield  {journal} {\bibinfo  {journal}
  {IEEE Transactions on Information Theory}\ }\textbf {\bibinfo {volume}
  {55}},\ \bibinfo {pages} {1898} (\bibinfo {year} {2009})},\ \bibinfo {note}
  {arXiv:quant-ph/0506189}\BibitemShut {NoStop}%
\bibitem [{\citenamefont {Uhlmann}(1976)}]{Uhl76}%
  \BibitemOpen
  \bibfield  {author} {\bibinfo {author} {\bibfnamefont {A.}~\bibnamefont
  {Uhlmann}},\ }\href@noop {} {\bibfield  {journal} {\bibinfo  {journal}
  {Reports on Mathematical Physics}\ }\textbf {\bibinfo {volume} {9}},\
  \bibinfo {pages} {273} (\bibinfo {year} {1976})}\BibitemShut {NoStop}%
\bibitem [{\citenamefont {Guha}(2008)}]{G08thesis}%
  \BibitemOpen
  \bibfield  {author} {\bibinfo {author} {\bibfnamefont {S.}~\bibnamefont
  {Guha}},\ }\emph {\bibinfo {title} {Multiple-User Quantum Information Theory
  for Optical Communication Channels}},\ \href@noop {} {Ph.D. thesis},\
  \bibinfo  {school} {Massachusetts Institute of Technology} (\bibinfo {year}
  {2008})\BibitemShut {NoStop}%
\bibitem [{\citenamefont {Wilde}(2016)}]{Wil15book}%
  \BibitemOpen
  \bibfield  {author} {\bibinfo {author} {\bibfnamefont {M.~M.}\ \bibnamefont
  {Wilde}},\ }\href@noop {} {\emph {\bibinfo {title} {From Classical to Quantum
  Shannon Theory}}}\ (\bibinfo {year} {2016})\ \bibinfo {note}
  {arXiv:1106.1445v7}\BibitemShut {NoStop}%
\bibitem [{\citenamefont {Reck}\ \emph {et~al.}(1994)\citenamefont {Reck},
  \citenamefont {Zeilinger}, \citenamefont {Bernstein},\ and\ \citenamefont
  {Bertani}}]{RZBB94}%
  \BibitemOpen
  \bibfield  {author} {\bibinfo {author} {\bibfnamefont {M.}~\bibnamefont
  {Reck}}, \bibinfo {author} {\bibfnamefont {A.}~\bibnamefont {Zeilinger}},
  \bibinfo {author} {\bibfnamefont {H.~J.}\ \bibnamefont {Bernstein}}, \ and\
  \bibinfo {author} {\bibfnamefont {P.}~\bibnamefont {Bertani}},\ }\href
  {\doibase 10.1103/PhysRevLett.73.58} {\bibfield  {journal} {\bibinfo
  {journal} {Physical Review Letters}\ }\textbf {\bibinfo {volume} {73}},\
  \bibinfo {pages} {58} (\bibinfo {year} {1994})}\BibitemShut {NoStop}%
\bibitem [{\citenamefont {Brandao}\ and\ \citenamefont {Datta}(2011)}]{BD11}%
  \BibitemOpen
  \bibfield  {author} {\bibinfo {author} {\bibfnamefont {F.~G. S.~L.}\
  \bibnamefont {Brandao}}\ and\ \bibinfo {author} {\bibfnamefont
  {N.}~\bibnamefont {Datta}},\ }\href {\doibase 10.1109/TIT.2011.2104531}
  {\bibfield  {journal} {\bibinfo  {journal} {IEEE Transactions on Information
  Theory}\ }\textbf {\bibinfo {volume} {57}},\ \bibinfo {pages} {1754}
  (\bibinfo {year} {2011})},\ \bibinfo {note} {arXiv:0905.2673}\BibitemShut
  {NoStop}%
\bibitem [{\citenamefont {Vedral}\ and\ \citenamefont {Plenio}(1998)}]{VP98}%
  \BibitemOpen
  \bibfield  {author} {\bibinfo {author} {\bibfnamefont {V.}~\bibnamefont
  {Vedral}}\ and\ \bibinfo {author} {\bibfnamefont {M.~B.}\ \bibnamefont
  {Plenio}},\ }\href@noop {} {\bibfield  {journal} {\bibinfo  {journal}
  {Physical Review A}\ }\textbf {\bibinfo {volume} {57}},\ \bibinfo {pages}
  {1619} (\bibinfo {year} {1998})},\ \bibinfo {note}
  {arXiv:quant-ph/9707035}\BibitemShut {NoStop}%
\bibitem [{\citenamefont {Hiai}\ and\ \citenamefont {Petz}(1991)}]{HP91}%
  \BibitemOpen
  \bibfield  {author} {\bibinfo {author} {\bibfnamefont {F.}~\bibnamefont
  {Hiai}}\ and\ \bibinfo {author} {\bibfnamefont {D.}~\bibnamefont {Petz}},\
  }\href@noop {} {\bibfield  {journal} {\bibinfo  {journal} {Communications in
  Mathematical Physics}\ }\textbf {\bibinfo {volume} {143}},\ \bibinfo {pages}
  {99} (\bibinfo {year} {1991})}\BibitemShut {NoStop}%
\bibitem [{\citenamefont {Buscemi}\ and\ \citenamefont {Datta}(2010)}]{BD10}%
  \BibitemOpen
  \bibfield  {author} {\bibinfo {author} {\bibfnamefont {F.}~\bibnamefont
  {Buscemi}}\ and\ \bibinfo {author} {\bibfnamefont {N.}~\bibnamefont
  {Datta}},\ }\href {\doibase 10.1109/TIT.2009.2039166} {\bibfield  {journal}
  {\bibinfo  {journal} {IEEE Transactions on Information Theory}\ }\textbf
  {\bibinfo {volume} {56}},\ \bibinfo {pages} {1447} (\bibinfo {year}
  {2010})},\ \bibinfo {note} {arXiv:0902.0158}\BibitemShut {NoStop}%
\bibitem [{\citenamefont {Wang}\ and\ \citenamefont {Renner}(2012)}]{WR12}%
  \BibitemOpen
  \bibfield  {author} {\bibinfo {author} {\bibfnamefont {L.}~\bibnamefont
  {Wang}}\ and\ \bibinfo {author} {\bibfnamefont {R.}~\bibnamefont {Renner}},\
  }\href {\doibase 10.1103/PhysRevLett.108.200501} {\bibfield  {journal}
  {\bibinfo  {journal} {Physical Review Letters}\ }\textbf {\bibinfo {volume}
  {108}},\ \bibinfo {pages} {200501} (\bibinfo {year} {2012})},\ \bibinfo
  {note} {arXiv:1007.5456}\BibitemShut {NoStop}%
\bibitem [{\citenamefont {M\"{u}ller-Hermes}(2012)}]{Mul12}%
  \BibitemOpen
  \bibfield  {author} {\bibinfo {author} {\bibfnamefont {A.}~\bibnamefont
  {M\"{u}ller-Hermes}},\ }\emph {\bibinfo {title} {Transposition in Quantum
  Information Theory}},\ \href@noop {} {Master's thesis} (\bibinfo {year}
  {2012})\BibitemShut {NoStop}%
\bibitem [{\citenamefont {Niset}\ \emph {et~al.}(2009)\citenamefont {Niset},
  \citenamefont {Fiur{\'{a}}{\v{s}}ek},\ and\ \citenamefont {Cerf}}]{NFC09}%
  \BibitemOpen
  \bibfield  {author} {\bibinfo {author} {\bibfnamefont {J.}~\bibnamefont
  {Niset}}, \bibinfo {author} {\bibfnamefont {J.}~\bibnamefont
  {Fiur{\'{a}}{\v{s}}ek}}, \ and\ \bibinfo {author} {\bibfnamefont
  {N.}~\bibnamefont {Cerf}},\ }\href {\doibase 10.1103/PhysRevLett.102.120501}
  {\bibfield  {journal} {\bibinfo  {journal} {Physical Review Letters}\
  }\textbf {\bibinfo {volume} {102}},\ \bibinfo {pages} {1} (\bibinfo {year}
  {2009})}\BibitemShut {NoStop}%
\bibitem [{\citenamefont {Braunstein}\ and\ \citenamefont
  {Kimble}(1998)}]{prl1998braunstein}%
  \BibitemOpen
  \bibfield  {author} {\bibinfo {author} {\bibfnamefont {S.~L.}\ \bibnamefont
  {Braunstein}}\ and\ \bibinfo {author} {\bibfnamefont {H.~J.}\ \bibnamefont
  {Kimble}},\ }\href@noop {} {\bibfield  {journal} {\bibinfo  {journal}
  {Physical Review Letters}\ }\textbf {\bibinfo {volume} {80}},\ \bibinfo
  {pages} {869} (\bibinfo {year} {1998})}\BibitemShut {NoStop}%
\bibitem [{Note2()}]{Note2}%
  \BibitemOpen
  \bibinfo {note} {In this experiment, the multiple access configuration is
  chosen to reduce the number of single photon detectors, which is the most
  expensive component in their protocol, where the simultaneous point-to-point
  protocol does not work.}\BibitemShut {Stop}%
\bibitem [{\citenamefont {Garc{\'{i}}a-Patr{\'{o}}n}\ and\ \citenamefont
  {Cerf}(2009)}]{GC09}%
  \BibitemOpen
  \bibfield  {author} {\bibinfo {author} {\bibfnamefont {R.}~\bibnamefont
  {Garc{\'{i}}a-Patr{\'{o}}n}}\ and\ \bibinfo {author} {\bibfnamefont
  {N.}~\bibnamefont {Cerf}},\ }\href {\doibase 10.1103/PhysRevLett.102.130501}
  {\bibfield  {journal} {\bibinfo  {journal} {Physical Review Letters}\
  }\textbf {\bibinfo {volume} {102}},\ \bibinfo {pages} {130501} (\bibinfo
  {year} {2009})}\BibitemShut {NoStop}%
\bibitem [{\citenamefont {Madsen}\ \emph {et~al.}(2012)\citenamefont {Madsen},
  \citenamefont {Usenko}, \citenamefont {Lassen}, \citenamefont {Filip},\ and\
  \citenamefont {Andersen}}]{MULFA12}%
  \BibitemOpen
  \bibfield  {author} {\bibinfo {author} {\bibfnamefont {L.~S.}\ \bibnamefont
  {Madsen}}, \bibinfo {author} {\bibfnamefont {V.~C.}\ \bibnamefont {Usenko}},
  \bibinfo {author} {\bibfnamefont {M.}~\bibnamefont {Lassen}}, \bibinfo
  {author} {\bibfnamefont {R.}~\bibnamefont {Filip}}, \ and\ \bibinfo {author}
  {\bibfnamefont {U.~L.}\ \bibnamefont {Andersen}},\ }\href {\doibase
  10.1038/ncomms2097} {\bibfield  {journal} {\bibinfo  {journal} {Nature
  communications}\ }\textbf {\bibinfo {volume} {3}},\ \bibinfo {pages} {1083}
  (\bibinfo {year} {2012})}\BibitemShut {NoStop}%
\bibitem [{\citenamefont {Takeoka}\ \emph {et~al.}(2015)\citenamefont
  {Takeoka}, \citenamefont {Jin},\ and\ \citenamefont {Sasaki}}]{TJS15}%
  \BibitemOpen
  \bibfield  {author} {\bibinfo {author} {\bibfnamefont {M.}~\bibnamefont
  {Takeoka}}, \bibinfo {author} {\bibfnamefont {R.-b.}\ \bibnamefont {Jin}}, \
  and\ \bibinfo {author} {\bibfnamefont {M.}~\bibnamefont {Sasaki}},\ }\href
  {\doibase 10.1088/1367-2630/17/4/043030} {\bibfield  {journal} {\bibinfo
  {journal} {New Journal of Physics}\ }\textbf {\bibinfo {volume} {17}},\
  \bibinfo {pages} {43030} (\bibinfo {year} {2015})}\BibitemShut {NoStop}%
\end{thebibliography}%

\appendix
\section{Supplementary Material 1: Calculation of $E^{\varepsilon}_R(\mathcal{T};A \overline{\mathcal{T}})_\phi$}

In this section we give more details regarding the calculation of the upper bound on  
the $\varepsilon$-relative entropy of entanglement (REE) 
$E^\varepsilon_R(\mathcal{T}^n;A^n \overline{\mathcal{T}}^n)_{\phi^{\otimes n}}$ where 
$\phi_{AB_1 \dots B_m} = \mathcal{L}_{A' \to B_1 \dots B_m} 
(|\Psi(N_S)\rangle \langle \Psi(N_S) | _{AA'})$ and 
$|\Psi(N_S)\rangle$ is a two-mode squeezed vacuum (TMVS) with 
average photon number $N_S$. 
$\mathcal{T}$ is some non-empty subset of $\{B_1, \dots, B_m\}$ 
and $\overline{\mathcal{T}}$ is its complement.  

\begin{figure}
\begin{center}
\includegraphics[width=90mm]{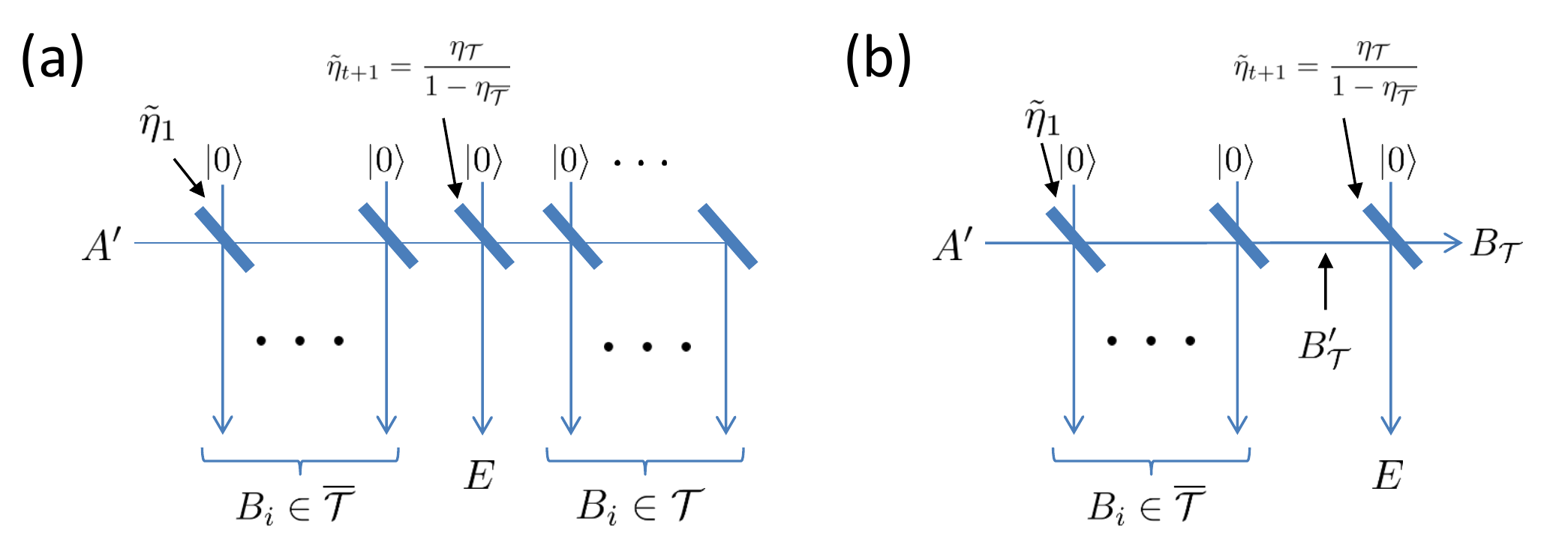}   %
\caption{\label{1-to-mQBC_reconfigured}
(a) Reconfiguration of the 1-to-$m$ pure-loss 
bosonic broadcast channel and 
(b) its unitary equivalent picture with respect to 
the bipartition $\mathcal{T};A\overline{\mathcal{T}}$. 
}
\end{center}
\end{figure}

To proceed with the calculation, it is critical to see that one can reconfigure 
the channel model described in Fig.~1(c) of the main text. 
Recall that the channel has $m+1$ transmittances 
$\eta_{B_1}, \cdots , \eta_{B_m}$, and 
$\eta_{E} \equiv 1-\sum_{i} \eta_{B_i}$. 
We can order these transmittances in some sequence and label it as 
$\eta_1, \eta_2, \cdots , \eta_m, \eta_{m+1}$.
Then for any ordering, it is possible to describe the channel 
by a sequence of $m$ beam splitters where the $j$-th beam splitter's 
transmittance is given by 
%%%%%%%%%%%%%%%%%%%%%%%%%%%
\begin{equation}
\label{eq:eta_i}
\tilde{\eta}_j = \frac{1-\sum_{k=1}^{j} \eta_k}{
1-\sum_{l=1}^{j-1} \eta_l} .
\end{equation}
%%%%%%%%%%%%%%%%%%%%%%%%%%%
Now, for each given $\mathcal{T}$ involving $t$ parties, 
consider the following specific ordering. 
For $\eta_i$ with $1 \le i \le t$, 
assign $\eta_{B_j}$ with $B_j \in \overline{\mathcal{T}}$, 
$\eta_{t+1} = \eta_E$, and for $\eta_i$ with $i > t+1$, assign 
$\eta_{B_j}$ with $B_j \in \mathcal{T}$.
Then the transmittance of the $t+1$ beam splitter is 
$\tilde{\eta}_{t+1} = \eta_{\mathcal{T}}/(1-\eta_{\overline{\mathcal{T}}})$ 
where $\eta_{\mathcal{T}} = \sum_{B_i \in \mathcal{T}} \eta_{B_i}$ 
and $\eta_{\overline{\mathcal{T}}} 
= \sum_{B_i \in \overline{\mathcal{T}}} \eta_{B_i}$
(see Fig.~\ref{1-to-mQBC_reconfigured}(a)).

%%%%%%%%%%%%%%%%%%%%%%%%%%%%%%%%%%%%%%%%%%%%
%%%%%%%%%%%%%%%%%%%%%%%%%%%%%%%%%%%%%%%%%%%%

In Fig.~\ref{1-to-mQBC_reconfigured}(a), 
the parties $A\mathcal{T}$ and $\overline{\mathcal{T}}$ 
are separated by a pure-loss 
channel with transmittance 
$\eta_\mathcal{T}/(1-\eta_{\overline{\mathcal{T}}})$. 
To simplify the picture, we consider the local unitary operations 
at each party, $A\mathcal{T}$ and $\overline{\mathcal{T}}$ 
(note that such unitaries do not change 
$E^\varepsilon_R(\mathcal{T}^n;A^n \overline{\mathcal{T}}^n)_{\phi^{\otimes n}}$). 
On $\mathcal{T}$, just undo all the beam splitters. 
Then as shown in Fig.~\ref{1-to-mQBC_reconfigured}(b), 
the system is described by $B_\mathcal{T}$, which is the output 
sent from $A$, and tensor products of vacuum that can be ignored 
in the rest of the calculation.

For the other side, $A\mathcal{T}$, 
we use the fact that our input state from the sender is a TMSV. 
Let $B'_{\mathcal{T}}$ denote the mode before the ($t+1$)-th beam splitter 
and let $\phi'_{A\overline{\mathcal{T}}B'_{\mathcal{T}}}$ 
be the TMSV in which only the first $t$ beam splitters are applied 
(see Fig.~\ref{1-to-mQBC_reconfigured}(b)). 
Observe that it is a pure state and its marginal $\phi'_{B'_{\mathcal{T}}}$ is 
a thermal state with average photon number 
$(1-\eta_{\mathcal{\overline{T}}})N_S$. 
Combining these two observations, we can conclude that 
the state has the following Schmidt decomposition: 
%%%%%%%%%%%%%%%%%%%%%%%%%%%
\begin{equation}
\label{eq:Schmidt_decomposition_m_receivers}
|\phi''\rangle_{A\overline{\mathcal{T}}B'_{\mathcal{T}}} 
= \sum_{m=0}^\infty \sqrt{\lambda_m((1-\eta_{\overline{\mathcal{T}}})N_S)} 
|\varphi_m\rangle_{A\overline{\mathcal{T}}} 
|m\rangle_{B'_{\mathcal{T}}} ,
\end{equation}
%%%%%%%%%%%%%%%%%%%%%%%%%%%
where $\lambda_m(N)$ is a thermal distribution, 
$|m\rangle$ is a photon number state, and 
$\{|\varphi_m\rangle_{A\overline{\mathcal{T}}}\}_m$ 
is some orthonormal basis. 
Since $\{|\varphi_m\rangle_{A\overline{\mathcal{T}}}\}_m$ 
is an orthonormal set, there exists a local unitary operation acting on 
$A\overline{\mathcal{T}}$ such that 
%%%%%%%%%%%%%%%%%%%%%%%%%%%
\begin{equation}
\label{eq:local_unitary}
U_{A\overline{\mathcal{T}}}: 
\, |\varphi_m\rangle_{A\overline{\mathcal{T}}} 
\to |m\rangle_A |{\rm aux}\rangle_{\overline{\mathcal{T}}} ,
\end{equation}
%%%%%%%%%%%%%%%%%%%%%%%%%%%
where $|{\rm aux}\rangle$ is some constant auxiliary state. Then we have 
%%%%%%%%%%%%%%%%%%%%%%%%%%%
\begin{align}
\label{eq:local_unitary_operation}
U_{A\overline{T}} 
|\phi''\rangle_{A\overline{\mathcal{T}}B'_{\mathcal{T}}} 
& =  |{\rm aux}\rangle_{\overline{\mathcal{T}}} 
|\Psi((1-\eta_{\overline{\mathcal{T}}})N_S)\rangle_{AB'_{\mathcal{T}}} 
\nonumber\\ & \equiv  
|\phi'''\rangle_{A\overline{\mathcal{T}}B'_{\mathcal{T}}}, 
\end{align}
%%%%%%%%%%%%%%%%%%%%%%%%%%%
where $|\Psi((1-\eta_{\overline{\mathcal{T}}})N_S)\rangle$ is 
a TMSV with average photon number $(1-\eta_{\overline{\mathcal{T}}})N_S$.

Let $\tilde{\phi}_{A\overline{\mathcal{T}}B_{\mathcal{T}}} 
= \mathcal{L}^{\tilde{\eta}_{t+1}}_{B'_\mathcal{T} \to B_\mathcal{T}} 
(|\phi'''\rangle
\langle\phi'''|_{A\overline{\mathcal{T}}B'_{\mathcal{T}}})$. 
Note that 
$\phi_{A\overline{\mathcal{T}}B_{\mathcal{T}}}$ equals 
$\mathcal{L}^{\tilde{\eta}_{t+1}}_{B'_\mathcal{T} \to B_\mathcal{T}} 
(|\phi''\rangle
\langle \phi'' |_{A\overline{\mathcal{T}}B'_{\mathcal{T}}})$ 
followed by $m-t-1$ beam splitters that have been already removed 
at the step from Fig.~\ref{1-to-mQBC_reconfigured}(a) 
to Fig.~\ref{1-to-mQBC_reconfigured}(b). 
Since these local unitary operations 
($U_{A\overline{\mathcal{T}}}$ and $m-t-1$ beam splitters) do not change 
the $\varepsilon$-REE between $A\overline{\mathcal{T}}$ and $\mathcal{T}$, we have 
$E^\varepsilon_R(B_\mathcal{T}^n;A^n\overline{\mathcal{T}}^n)_{\tilde{\phi}^{\otimes n}} 
= E^{\varepsilon}_R(\mathcal{T}^n;A^n\overline{\mathcal{T}}^n)_{\phi^{\otimes n}}$
Moreover, $E^{\varepsilon}_R(B_\mathcal{T}^n;A^n\overline{\mathcal{T}}^n)_{\tilde{\phi}^{\otimes n}}$ 
is the $\varepsilon$-REE for the TMSV with $(1-\eta_{\overline{\mathcal{T}}})N_S$ 
followed by 
$\mathcal{L}^{\tilde{\eta}_{t+1}}_{B'_\mathcal{T} \to B_\mathcal{T}}$ where 
$\tilde{\eta}_{t+1} = \eta_\mathcal{T}/(1-\eta_{\overline{\mathcal{T}}})$. 
Then by using the result in \cite{WTB17}, 
the REE bound for $N_S \to \infty$ turns out to be 
%%%%%%%%%%%%%%%%%%%%%%%%%%%
\begin{align}
\label{eq:REE_m-receiver}
\sum_{B_i \in \mathcal{T}} (E_{AB_i} + K_{AB_i}) &
\le 
\frac{1}{n} E^{\varepsilon(N_S)}_R (\mathcal{T}^n ; A^n \overline{\mathcal{T}}^n)_{\phi^{\otimes n}}
\nonumber \\ & \leq  
\log_2 \left( \frac{1}{
1-\frac{\eta_{\mathcal{T}}}{1-\eta_{\overline{\mathcal{T}}}}}
\right) + C(\varepsilon)/n
\nonumber \\ & =  
\log_2 \left( \frac{1-\eta_{\overline{\mathcal{T}}}}{
1-\eta_{\mathcal{T}} - \eta_{\overline{\mathcal{T}}}} \right) + C(\varepsilon)/n
\nonumber \\ & =  
\log_2 \left( \frac{1-\eta_{\overline{\mathcal{T}}}}{
1-\eta_{\mathcal{B}}} \right) + C(\varepsilon)/n,
\end{align}
%%%%%%%%%%%%%%%%%%%%%%%%%%%
where 
$\eta_{\mathcal{B}} = \eta_{\mathcal{T}} + \eta_{\overline{\mathcal{T}}}$.

\section{Supplementary Material 2: The rate sum calculation for the 1-to-m symmetric pure-loss 
broadcast channel}

Here we calculate the achievable rate sum in the 1-to-m symmetric 
pure-loss channel with equal transmittance $\eta/m \equiv \tilde{\eta}$ to each receiver. 
We restrict the distillation scenario such that all receivers achieve the same rate.  
Though only the entanglement distillation is considered in the main text, the same achievable rate sum 
holds for secret key generation. Thus we here treat both entanglement and secret key. 
Let $E_{\rm sum}=\max \sum_i^m E_{AB_i}$ and 
$K_{\rm sum}=\max \sum_i^m K_{AB_i}$ be the rate sum of the entanglement and secret key distillations, respectively.
Then we show that the maximum achievable $E_{\rm sum}+K_{\rm sum}$ 
is given by
%%%%%%%%%%%%%%%%%%%%%%%%%%%
\begin{equation}
\label{eq:maximum_sum_rate}
E_{\rm sum}+K_{\rm sum} = -\log_2 (1-\eta).  
\end{equation}
%%%%%%%%%%%%%%%%%%%%%%%%%%%
whereas the achievable rate sum for the time sharing strategy is bounded by
%%%%%%%%%%%%%%%%%%%%%%%%%%%
\begin{equation}
\label{eq:time_sharing_rate_sum}
E_{\rm sum}^{\rm ts}+K_{\rm sum}^{\rm ts} \le -\log_2 (1-\eta/m). 
\end{equation}
%%%%%%%%%%%%%%%%%%%%%%%%%%%

Let us first consider the time-sharing bound.
The point-to-point capacity for the sender and each receiver is given by $-\log_2 (1-\eta/m)$. 
To achieve the same rate for each receiver, the receiver should time share the channel equally.
Then each receiver achieves the rate of $- \frac{1}{m} \log_2 (1-\eta/m)$ and sum up for the $m$ receivers, 
the rate sum is given by $- \log_2 (1-\eta/m)$.

For the maximum achievable rate sum, consider the capacity region for the symmetric pure-loss channel. 
According to Theorem 1, the capacity region for a pure-loss symmetric QBC is given by a set of inequalities: 
%%%%%%%%%%%%%%%%%%%%%%%%%%%
\begin{equation}
\label{eq:capacity_region_symmetric_QBC}
\sum_{B_i \in \mathcal{T}} E_{AB_i} + K_{AB_i} \le \log_2 \frac{1-(m-|\mathcal{T}|) \tilde{\eta}}{1-m\tilde{\eta}}. 
\end{equation}
%%%%%%%%%%%%%%%%%%%%%%%%%%%
Applying the restriction that all the receivers achieve the same rate $E_{AB}$ and $K_{AB}$, 
the above rate region expression is reduced to  
%%%%%%%%%%%%%%%%%%%%%%%%%%%
\begin{equation}
%\label{eq:capacity_symmetric_QBC}
E_{AB} + K_{AB} \le \frac{1}{l} 
\log_2 \frac{1-(m-l) \tilde{\eta}}{1-m\tilde{\eta}}, 
\end{equation}
%%%%%%%%%%%%%%%%%%%%%%%%%%%
for all integers $l$ satisfying $ 1 \le l \le m$. 
Thus the maximum rate sum is bounded by 
%%%%%%%%%%%%%%%%%%%%%%%%%%%
\begin{equation}
\label{eq:capacity_symmetric_QBC}
E_{\rm sum} + K_{\rm sum} \le \frac{m}{l} 
\log_2 \frac{1-(m-l) \tilde{\eta}}{1-m\tilde{\eta}}, 
\end{equation}
%%%%%%%%%%%%%%%%%%%%%%%%%%%
for all $l$. 
Then to prove Eq.~(\ref{eq:maximum_sum_rate}), 
it is sufficient to show that the right hand side of 
(\ref{eq:capacity_symmetric_QBC}) is monotonically decreasing 
with respect to $l$. 
Let $f(x) \equiv (m/x) \log_2 [\{1-(m-x) \tilde{\eta}\}/\{1-m\tilde{\eta}\}]$ 
for real $x$ with $0 < x \le 1$. 
Then 
%%%%%%%%%%%%%%%%%%%%%%%%%%%
\begin{equation}
\label{eq:f'(x)}
f'(x) = \frac{m}{x \ln 2} \left( 
\frac{\tilde{\eta}}{1-(m-x)\tilde{\eta}} 
- \frac{1}{x} \ln \left[ \frac{1-(m-x) \tilde{\eta}}{1-m \tilde{\eta}} \right] \right) . 
\end{equation}
%%%%%%%%%%%%%%%%%%%%%%%%%%%
Since $\frac{1-(m-x) \tilde{\eta}}{1-m \tilde{\eta}} \equiv y > 1$, 
we have 
%%%%%%%%%%%%%%%%%%%%%%%%%%%
\begin{eqnarray}
\label{eq:log_y}
\ln y & = & 2 \left[ \frac{y-1}{y+1} 
+ \frac{1}{3} \left(\frac{y-1}{y+1}\right)^3 
+ \frac{1}{5} \left(\frac{y-1}{y+1}\right)^5 + \cdots \right]
\nonumber\\ & \ge & \frac{2(y-1)}{y+1} . 
\end{eqnarray}
%%%%%%%%%%%%%%%%%%%%%%%%%%%
Applying this into the second term of (\ref{eq:f'(x)}), we have 
%%%%%%%%%%%%%%%%%%%%%%%%%%%
\begin{eqnarray}
\label{eq:f'(x)_2}
f'(x) & \le & \frac{m}{x \ln 2} \left( 
\frac{\tilde{\eta}}{1-(m-x)\tilde{\eta}} 
- \frac{2\tilde{\eta}}{2(1-m\tilde{\eta}) + \tilde{\eta}x} 
\right) 
\nonumber\\ & = & 
- \frac{2 m \tilde{\eta}^2}{(\ln 2) (1-m\tilde{\eta}+x \tilde{\eta})
\{2(1-m\tilde{\eta})+x\tilde{\eta} \}}
\nonumber\\ & \le & 0.
\end{eqnarray}
%%%%%%%%%%%%%%%%%%%%%%%%%%%
Note that $m \tilde{\eta} = \eta \le 1$. 
Therefore, the bound (\ref{eq:capacity_symmetric_QBC}) is tightest at $l=m$ 
implying Eq.~(\ref{eq:maximum_sum_rate}). 

\section{Supplementary Material 3: Broadcast channel quantum key distribution}

In this section, we give details of the broadcast channel continuous variable quantum key distribution (BC-CVQKD). 
Our protocol is based on the CVQKD protocol with squeezed state proposed by Garc\'{i}a-Patr\'{o}n and Cerf (GC09) \cite{GC09}. 
Here we briefly review the GC09 protocol and then extend it to our BC-CVQKD. 

\begin{figure}
\begin{center}
\includegraphics[width=90mm]{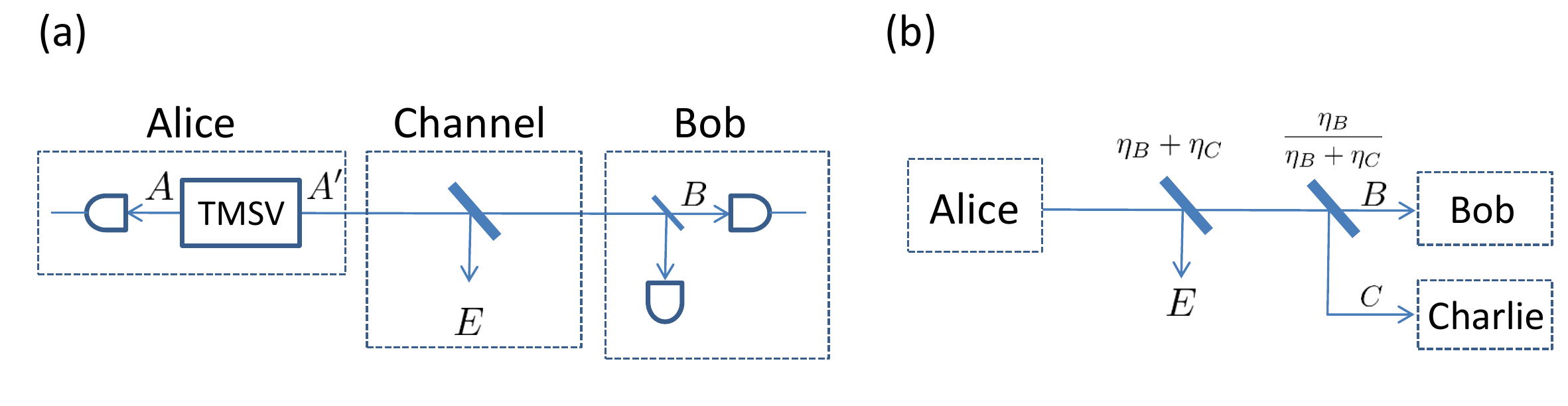}   %
\caption{\label{BC-CVQKD}
(a) CVQKD with squeezed state \cite{GC09} and   
(b) its' extension to the 1-to-2 broadcast-CVQKD.
}
\end{center}
\end{figure}

Figure \ref{BC-CVQKD}(a) shows the entanglement-based description of the GC09 protocol 
(in practice, it is implemented by a randomly modulated single-mode squeezed state). 
Alice prepares a two-mode squeezed vacuum (TMSV) state in modes $A$ and $A'$, 
where mode $A$ is measured by a homodyne detector with a random choice of the quadrature basis 
and mode $A'$ is sent to Bob through a point-to-point quantum channel, possibly attacked by Eve. 
Bob measures the received state by a heterodyne detector which simultaneously measures two quadratures. 
After the transmission of $n$ pulses of the TMSV, Alice announces her basis choice for each measurement to Bob
through a public classical channel. Bob uses the measurement outcomes for the corresponding bases to make 
the classical postprocessing, called reverse reconciliation (RR), in order to distill the secret key 
(thus Bob's measurement is effectively modeled by a 50\% pure-loss followed by a homodyne detector \cite{GC09}).  
Let $X$ and $Y$ be the measurement outcomes for Alice and Bob that are used to distill the key, 
and let $E$ be the quantum system held by Eve, representing the environment of the quantum channel. 
To simplify the setting, hereafter we consider a pure-loss bosonic channel as the quantum channel. 
The key generation rate between Alice and Bob is then given by 
%%%%%%%%%%%%%%%%%%%%%%%%%%%
\begin{equation}
\label{eq:GC09_key_rate}
K_{AB}  = I(X;Y) - I(Y;E)_\rho .
\end{equation}
%%%%%%%%%%%%%%%%%%%%%%%%%%%

%%%%%%%%%%%%%%%%%%%%%%%%%%%%%%%%%%%%%%%%%%%%
%%%%%%%%%%%%%%%%%%%%%%%%%%%%%%%%%%%%%%%%%%%%

We extend this GC09 protcol to the broadcast channel scenario. 
Figure \ref{BC-CVQKD}(b) depicts the simplest case where one sender, Alice, sends quantum signal to two receivers, 
Bob and Charlie, through a 1-to-2 pure-loss bosonic broadcast channel. 
The goal of the protocol is to share bipartite secret keys $K_{AB}$ (Alice-Bob) and $K_{AC}$ (Alice-Charlie).  
Here we consider a conservative scenario where the key is secure against any third parties. 
For example, to distill $K_{AB}$, the environment of the channel (Eve) and Charlie are both regarded as eavesdroppers. 

Let us first describe a simple operation of the point-to-point GC09 protocol.  
Alice prepares the TMSV in modes $A$ and $A'$, measures mode $A$ by a homodyne receiver and sends mode $A'$ 
to Bob and Charlie through a pure-loss broadcast channel. 
Bob and Charlie make heterodyne measurements. 
Then they separately do the key distillation with Alice. 
Since the key distillation is performed on classical data (measurement outcomes), 
the key distillation for Alice-Bob and Alice-Charlie can be performed for the same measurement outcomes. 
Let $X$, $Y$, and $Z$ be the measurement outcomes at Alice, Bob, and Charlie, to be used for the key distillation. 
Then they can achieve the key rate region
%%%%%%%%%%%%%%%%%%%%%%%%%%%
\begin{equation}
\label{eq:BC-CVQKD_naive}
\begin{array}{l}
K_{AB} \le I(X;Y) - I(Y;CE)_\rho , \\
K_{AC} \le I(X;Z) - I(Z;BE)_\rho .
\end{array}
\end{equation}
%%%%%%%%%%%%%%%%%%%%%%%%%%%

Now the question is if we can surpass the above rate region by an explicit QKD protocol. 
In the following, we describe the protocol where Charlie cooperate to Alice and Bob to distill $K_{AB}$ and 
Bob also helps to distill $K_{AC}$, which we call broadcast-CVQKD (BC-CVQKD) protocol. 
Again Alice sends one part of her TMSVs to Bob and Charlie and then they share $X$, $Y$, and $Z$. 
Then one of the receivers, for example Charlie, makes the standard point-to-point RR with Alice. 
After that they can achieve the secret key rate of $K_{AC} = I(X;Z) - I(Z;BE)_\rho$. 
Note that during this process, Alice reconstructs Charlie's information $Z$ and thus she holds $X$ and $Z$ 
as her classical information. 
Then Bob makes the RR with Alice where Alice can use $X$ and $Z$ as her classical information. 
Thus they can establish the key rate $K_{AB} = I(XZ;Y) - I(Y;CE)_\rho$. 
The key rate difference between this BC-CVQKD and the simple point-to-point protocol described above is 
$I(XZ;Y) - I(X;Y) = I(Y;Z|X) \ge 0$ 
where the inequality holds since the conditional mutual information is a non-negative function. 
Thus our BC-CVQKD protocol can always achieve better or at least the same performance as the simple point-to-point-QKD based protocol. 
In summary, the achievable rate pair is 
%%%%%%%%%%%%%%%%%%%%%%%%%%%
\begin{equation}
\label{eq:BC-CVQKD_rate_pair_1}
\begin{array}{l}
K_{AB}  = I(XZ;Y) - I(Y;CE)_\rho , \\
K_{AC} = I(X;Z) - I(Z;BE)_\rho .
\end{array}
\end{equation}
%%%%%%%%%%%%%%%%%%%%%%%%%%%
Similarly, when Bob first makes the RR, then they can achieve the rate pair
%%%%%%%%%%%%%%%%%%%%%%%%%%%
\begin{equation}
\label{eq:BC-CVQKD_rate_pair_2}
\begin{array}{l}
K_{AB}  = I(X;Y) - I(Y;CE)_\rho , \\
K_{AC} = I(XY;Z) - I(Z;BE)_\rho .
\end{array}
\end{equation}
%%%%%%%%%%%%%%%%%%%%%%%%%%%
Thus the achievable rate region of our protocol is given by the time sharing of 
(\ref{eq:BC-CVQKD_rate_pair_1}) and (\ref{eq:BC-CVQKD_rate_pair_2}). 

Each of the information quantities in (\ref{eq:BC-CVQKD_rate_pair_1}) and (\ref{eq:BC-CVQKD_rate_pair_2}) are given 
as follows. 
The classical mutual informations are derived by calculating the Shannon entropies, 
$H(X)$, $H(Y)$, $H(Z)$, $H(XY)$, $H(XZ)$, and $H(XYZ)$. 
Since TMSV and pure-loss channel are Gaussian state and channel, respectively, we use the covariance matrix 
approach  (here we basically use the definition in \cite{TJS15}). 
Let $\mu$ be the average photon number of the TMSV and $v=2\mu+1$. 
The quantum state held by Alice, Bob, and Charlie before their measurements is described by the covariance matrix 
%%%%%%%%%%%%%%%%%%%%%%%%%%%
\begin{equation}
\label{eq:gamma_AB}
\gamma_{ABC} = 
\left[
\begin{array}{cccccc}
a & d & f & 0  & 0 & 0 \\
d & b & e & 0 & 0 & 0 \\
f & e & c & 0 & 0 & 0 \\
0 & 0 & 0 & a & -d & -f \\
0 & 0 & 0 & -d & b & e \\
0 & 0 & 0 & -f & e & c \\
\end{array}
\right] ,
\end{equation}
%%%%%%%%%%%%%%%%%%%%%%%%%%%
where $a = v$, $b= \frac{1}{2} \eta_B (v-1)  + 1$, $b= \frac{1}{2} \eta_C (v-1) + 1$, 
$d = \sqrt{\frac{1}{2}\eta_B (v^2-1) }$, $e = \frac{1}{2}\sqrt{\eta_B \eta_C} (v-1)$, and 
$f = \sqrt{\frac{1}{2}\eta_B (v^2-1)}$.  
Note that the effective 50\% loss at the heterodyne measurements are already included. 
The covariance matrix for the homodyne measurements at Alice, Bob, and Charlie is given by 
%%%%%%%%%%%%%%%%%%%%%%%%%%%
\begin{equation}
\label{eq:gamma_M}
\gamma_{M} = 
\left[
\begin{array}{cc}
e^{-2r} I_3 & 0 \\
0 & e^{2r} I_3 \\
\end{array}
\right] ,
\end{equation}
%%%%%%%%%%%%%%%%%%%%%%%%%%%
where $I_3$ is a $3 \times 3$ identity matrix and $r \to \infty$. 
The probability density function after the measurement is given by 
%%%%%%%%%%%%%%%%%%%%%%%%%%%
\begin{eqnarray}
\label{eq:pdf_ABC}
P_{XYZ} (x,  y, z) & = & \frac{1}{\pi \sqrt{{\rm det} (\gamma_{ABC} + \gamma_M)}} 
\nonumber\\ && \times
\exp\left[
-{\bf x}^T \frac{1}{(\gamma_{ABC} + \gamma_M)} {\bf x} \right] ,
\end{eqnarray}
%%%%%%%%%%%%%%%%%%%%%%%%%%%
where ${\bf x} = [x, y, z]^T$ and ${\rm det} \gamma$ is the determinant of $\gamma$. 
Then we get the differential entropy of this as 
%%%%%%%%%%%%%%%%%%%%%%%%%%%
\begin{eqnarray}
\label{eq:H(XYZ)}
H(XYZ) & = & \frac{1}{2} \log_2 (\pi e)^3 {\rm det} \gamma_{ABC} 
\nonumber\\ & = & 
\frac{1}{2} \log_2 (\pi e)^3 \left\{ \left( 1-\frac{\eta_B+\eta_C}{2} \right) (v-1) +1 \right\} .
\nonumber\\
\end{eqnarray}
%%%%%%%%%%%%%%%%%%%%%%%%%%%
Since the covariance matrix of the marginal system is obtained by simply taking a corresponding submatrix of $\gamma_{ABC}$, 
we get the necessary differential entropies in a similar way as 
%%%%%%%%%%%%%%%%%%%%%%%%%%%
\begin{eqnarray}
\label{eq:entropies}
H(X) & = & \frac{1}{2} \log_2 (\pi e) v , \\
H(Y) & = & \frac{1}{2} \log_2 (\pi e) \left\{ \frac{\eta_B}{2} (v-1) +1 \right\}, \\
H(Z) & = & \frac{1}{2} \log_2 (\pi e) \left\{ \frac{\eta_C}{2} (v-1) +1 \right\}, \\
H(XY) & = & \frac{1}{2} \log_2 (\pi e)^2 \left\{ \left( 1-\frac{\eta_B}{2} \right) (v-1) +1 \right\}, \\
\label{eq:entropies_last}
H(XZ) & = & \frac{1}{2} \log_2 (\pi e)^2 \left\{ \left( 1-\frac{\eta_C}{2} \right) (v-1) +1 \right\}.
\end{eqnarray}
%%%%%%%%%%%%%%%%%%%%%%%%%%%

Below, we calculate the quantum mutual information $I(Y;CE)_\rho = H(CE)_\rho - H(CE|Y)_\rho$. 
Since $C$ and $E$ are regarded as one system, it is useful to rewrite the channel in Fig.~\ref{BC-CVQKD}(b) 
as the one in Fig.~\ref{BC-CVQKD}(c). 
Here after system $CE$ is sometimes denoted as $C'$ for simplicity.
The covariance matrix for the quantum system held by Charlie-Eve and Bob (after the 50\% loss in the heterodyne) is 
%%%%%%%%%%%%%%%%%%%%%%%%%%%
\begin{equation}
\label{eq:gamma_EB}
\gamma_{C'B} = 
\left[
\begin{array}{cccc}
a' & c' & 0 & 0 \\
c' & b' & 0 & 0 \\
0 & 0 & a' & -c' \\
0 & 0 & -c' & b' 
\end{array}
\right] ,
\end{equation}
%%%%%%%%%%%%%%%%%%%%%%%%%%%
where $a' = (1-\eta_B)(v-1) + 1$, $b' = \frac{\eta_B}{2} (v-1) + 1$, $c' = -\sqrt{\frac{\eta_B}{2} (1-\eta_B)} (v-1)$. 
Since the von Neumann entropy of a single-mode Gaussian state is given by $g((\lambda-1)/2)$, 
where $g(x) = (x+1) \log_2 (x+1) - x \log_2 x$ and $\lambda$ is a symplectic eigenvalue of the covariance matrix,  
we have 
%%%%%%%%%%%%%%%%%%%%%%%%%%%
\begin{equation}
\label{eq:H(E)}
H(CE)_\rho = g \left( \frac{(1-\eta_B)(v-1)}{2} \right) .
\end{equation}
%%%%%%%%%%%%%%%%%%%%%%%%%%%

Eve's quantum state conditioned on Bob's measurement is given as follows. 
By changing the order of the matrix entry, $\gamma_{C'B}$ is represented as 
%%%%%%%%%%%%%%%%%%%%%%%%%%%
\begin{equation}
\label{eq:gamma_example}
\gamma_{C'B} = \left[
\begin{array}{cc}
A & C \\
C^T & B \\
\end{array}
\right] ,
\end{equation}
%%%%%%%%%%%%%%%%%%%%%%%%%%%
where $A = a' I_2$, $B = b' I_2$ and $I_2$ is a $2\times2$ identity matrix, and 
%%%%%%%%%%%%%%%%%%%%%%%%%%%
\begin{equation}
C = \left[
\begin{array}{cc}
c' & 0 \\
0 & -c' 
\end{array}
\right] .
\end{equation}
%%%%%%%%%%%%%%%%%%%%%%%%%%%
Now we apply a homodyne measurement on Bob's state The covariance matrix for a single-mode homodyne measurement 
is given by 
%%%%%%%%%%%%%%%%%%%%%%%%%%%
\begin{equation}
\gamma_M = \left[
\begin{array}{cc}
e^{-2r} & 0 \\
0 & e^{2r} 
\end{array}
\right] ,
\end{equation}
%%%%%%%%%%%%%%%%%%%%%%%%%%%
with $r \to \infty$. 
Then Eve's state in $C'$ conditioned on Bob's homodyne measurement is described by 
%%%%%%%%%%%%%%%%%%%%%%%%%%%
\begin{equation}
\label{eq:gamma_example}
\gamma_{C'|y} = A - C^T \frac{1}{B + \gamma_M} C  = 
\left[
\begin{array}{cc}
\alpha & 0 \\
0 & \beta \\
\end{array}
\right] ,
\end{equation}
%%%%%%%%%%%%%%%%%%%%%%%%%%%
where  
%%%%%%%%%%%%%%%%%%%%%%%%%%%
\begin{equation}
\label{eq:alpha_beta}
\alpha = \frac{(1-\eta_B)(v-1)}{\frac{1}{2}\eta_B (v-1) +1} + 1 ,
\end{equation}
%%%%%%%%%%%%%%%%%%%%%%%%%%%
and $\beta = (1-\eta_B)(v-1)+1$. 
Note that $\gamma_{C'|y}$ is independent on the homodyne measurement outcome $y$, 
i.e. same for any $y$.  
As a consequence, we get the conditional quantum entropy as 
%%%%%%%%%%%%%%%%%%%%%%%%%%%
\begin{equation}
\label{eq:H(E|Y)}
H(CE|Y)_\rho = g \left( \frac{\sqrt{\alpha\beta} - 1}{2} \right)
\end{equation}
%%%%%%%%%%%%%%%%%%%%%%%%%%%
Combining (\ref{eq:H(XYZ)})--(\ref{eq:entropies_last}), (\ref{eq:H(E)}), and (\ref{eq:H(E|Y)}), one can compute 
(\ref{eq:BC-CVQKD_rate_pair_1}) and (\ref{eq:BC-CVQKD_rate_pair_2}).
The other mutual information, $I(Z;BE)_\rho$ is also obtained in a similar way.

\end{document}